-------------------------------------------------------------------

\documentstyle[12pt]{article}
\newcommand{\be}{\begin{equation}}
\newcommand{\ee}{\end{equation}}
\newcommand{\bea}{\begin{eqnarray}}
\newcommand{\eea}{\end{eqnarray}}

\newcommand{\average}[1]{\makebox{$\langle{#1}\rangle$}}

\newcommand{\Tr}{\mathop{\mbox{Tr}}\nolimits}
\newcommand{\tr}{\mathop{\mbox{tr}}\nolimits}

\newcommand{\Ima}{\mathop{\mbox{Im}}\nolimits}
\newcommand{\sumn}{\mathop{\sum_{n=0}^{\infty}}}
\newcommand{\sumal}{\mathop{\sum_{\alpha=1}^{g_n}}}
\newcommand{\sumo}{\mathop{\sum_{n\,\in\,\Omega}}}
\newcommand{\sumop}{\mathop{\sum_{n\,\in\,\Omega_+}}}
\newcommand{\sumom}{\mathop{\sum_{n\,\in\,\Omega_-}}}
\newcommand{\sumopm}{\mathop{\sum_{n\,\in\,\Omega_\pm}}}

\newcommand{\Psic}{{\overline \Psi}}
\begin{document}

\centerline{\bf SEMICLASSICAL EXPANSIONS UP TO $\hbar^4 $-ORDER}

\centerline{\bf IN RELATIVISTIC NUCLEAR  PHYSICS}
\vskip.5cm

\centerline{ J. Caro,${}^1$
E. Ruiz Arriola${}^{1,2}$ and L.L. Salcedo${}^1$  }
\vskip0.5cm
\centerline {${}^1$ Departamento de F{\'{\i}}sica Moderna, Universidad
de Granada}
\centerline {E-18071 Granada, Spain }
\vskip.5cm
\centerline { ${}^2$ National Institute for Nuclear Physics and High
Energy Physics (NIKHEF-K) }
\centerline {1009-DB Amsterdam, The Netherlands}
\vskip2cm

\centerline{\bf ABSTRACT}
\vskip1cm

We present the first calculation of the $\hbar^4$-Wigner--Kirkwood
corrections to a relativistic
system of fermions in the presence of external
scalar and vector potentials. The method we propose allows to compute
efficiently semiclassical corrections to one body operators such as
mean energies and local densities. It also preserves gauge invariance
and produces explicitly convergent results despite some apparent
divergencies at the classical turning points. As a byproduct we obtain
the $\hbar^4$ corrections stemming from the polarization of the Dirac
sea. We apply our results to the relativistic $\sigma$-$\omega$
Lagrangian in the Hartree valence approximation. We compare the
semiclassical expansion with
the exact result and with the Strutinsky average whenever it can be
obtained. We find that the $\hbar^4 $ corrections are much smaller than
typical shell effects. Our results provide convincing arguments to
neglect higher than second order $\hbar$ effects in the Wigner--Kirkwood
scheme to relativistic nuclear physics in the mean field approximation.
\vskip1cm
\centerline{\sl October 1994 }
\vskip2cm\hbox{UG-DFM-32/94 \hfill}

\newpage

\section{Introduction }

Relativistic mean field theories for nuclear systems have already a long
history \cite{Wa74} and have found a wealth of applications
\cite{Se86,Re89, Se92}. They use as a
starting point a relativistic quantum field Lagrangian which hopefully
includes the relevant degrees of freedom, i.e. nucleons and several
mesons. In the Hartree approximation it is possible to reproduce the
saturation properties for nuclear matter and the shell structure of
single particle states is accurately described due to the correct sign
and size of the spin-orbit interaction. Many of the calculations are performed
in practice for spherical closed shell nuclei, where shell effects are
expected to be least significant. For these systems many authors have
proposed to employ a relativistic Thomas-Fermi (TF) model \cite{Se86}
and generalizations thereof
\cite{Ce90,We91,Sp92,Vo92,Ce92,Ce93a,Ce93b,Ru93,Ce94}.

{}From a purely computational point of view, the study of the
semiclassical expansion for relativistic systems is far
from being a trivial generalization of well-known non-relativistic
methods which have been discussed at length for nuclear systems in the
literature (for reviews see e.g. \cite{Ri80,Br85} and references
therein). The main problem is that whereas in the Schr\"odinger
equation the Planck's constant $\hbar$ appears in the second power, in
the Dirac equation
$\hbar$ appears in the first power. The final results for bulk
properties do not depend, however, on the sign of $\hbar$ due to parity
invariance. This simple
circumstance makes the direct generalization of non-relativistic methods
such as e.g. Wigner transformation or others quite time consuming and
generate extremely complicated expressions in intermediate stages of
the calculation. An additional difficulty in the relativistic case is
the appearance of four dimensional spinor matrices. One should also say
that the treatment of nuclei requires at least a scalar potential in
addition to the usual electrostatic potential of atomic physics,
therefore making the computations much more involved. This
probably explains why until now only the
corrections up to $\hbar^2 $ have been computed in relativistic nuclear
systems \cite{Ce90,We91} and up to $\hbar^4 $ in atomic systems
\cite{Po91}. In this respect one should mention that in the
non-relativistic case the $\hbar^4$, and $\hbar^6$ have been known
since a long time \cite{Ho73,Mu81} and quite recently
$\hbar^8 $-order corrections have been determined
\cite{Sh92}.
The question arises whether there are reasons to neglect higher
orders than $\hbar^2 $, besides the desire to keep the problem
at a manageable level.

The answer to this question cannot be given in a closed form, because
in general it is
not known whether the semiclassical series is convergent or
only asymptotic. In the absence of exact theorems only numerical
investigation remains. As a practical rule there are cases like e.g.
the non-relativistic harmonic oscillator where the Strutinsky average
is known to reproduce
the complete Wigner-Kirkwood (WK) result \cite{Br73}.
This rule seems also to be
fulfilled for the relativistic harmonic potential \cite{Ce93b} in order
WK$\hbar^2$.\footnote{The notation emphasizes that we are performing
a WK expansion throughout, rather than an extended TF, that is, we expand
in gradients of the potentials instead of gradients of the densities.}
However, in the cases
of relevance for nuclear physics, i.e. short range and finite depth
potentials, the Strutinsky average is in general not well defined,
although nothing prevents from an order by order semiclassical
calculation.

In the present paper we propose an efficient computational method which
allows to compute up to WK$\hbar^4 $-corrections of the semiclassical
expansion for one-body operators following the method proposed in
previous work \cite{Ru93}. This method is fully equivalent
to the standard WK expansion \cite{Wi32, Wi34, Ki33} for
one body operators; the main advantage is that one does not need to
evaluate non-diagonal elements of the Green's functions and traces back
all physical quantities to the cumulative number of states.
As an aside we compute analytically the WK$\hbar^4$-corrections coming
from the polarization of the Dirac sea. We apply our results to the
relativistic $\sigma$-$\omega$ model in the Hartree valence
approximation and make a numerical estimate of the
WK$\hbar^ 4 $-corrections to the fermionic contribution to the total
energy. To set in the proper scales we adjust the
parameters to reproduce the nuclear matter saturation properties and the
charge mean squared radius of ${}^{40}$Ca \cite{Ho81b}. In doing so we
do not pretend to make a full account of finite nuclei properties, since
there are more realistic generalizations of this model to include
Coulomb repulsion, tensor force, etc. \cite{Se86,Re89,Se92}.
Nevertheless we feel that the general features obtained in the present
study should not be substantially modified by these extensions of the
$\sigma$-$\omega$ model.

The outline of the paper is as follows. In section 2 we review the
properties of a $D$-dimensional non-relativistic case to
illustrate the basic mathematical objects such as the determinant of
the Schr\"odinger operator and our method to perform
semiclassical expansions in the simplest situation. As an example we
apply our formalism to the three dimensional harmonic oscillator.
We also discuss how a reordering of the semiclassical series proposed
recently \cite{Sc93,Kr90} can be incorporated into the formalism, proving
extremely convenient for numerical computations.
Section 3 studies the  relativistic case and the particular features of
the determinant of the Dirac equation and explicit calculations up to
$\hbar^4 $-order are also presented. As an illustration we
apply and discuss those corrections in the analytically solvable example
of relativistic harmonic potentials with special emphasis on the
distributional character of the quantities involved. We also show an
explicit calculation of the (finite) WK$\hbar^4$-corrections arising from
the polarization of the Dirac sea. In section  4 we quickly review the
$\sigma$-$\omega$ model together with the parameter fixing and give
numerical estimates to the WK$\hbar^4 $-corrections to the Hartree energy.
We also provide a prescription how to avoid the distributions in
practical calculations. Finally, in section 5 we draw our conclusions
and present some perspectives for future research.

\newpage

\section{ The Number of States Formalism}
\label{non-rel}

\subsection{\sl Cumulative Number of States}

In this work we will consider objects which involve single
sums, i.e. one body operators. All the information concerning them
can be deduced from the single particle level density. As it will become
clear below, it is more convenient to deal with the cumulative number
of states. Let us consider the eigenvalue problem
\be
H\,\psi_{n\alpha}(x) = E_n \, \psi_{n\alpha}(x)
\ee
where $H$ is the single particle Hamiltonian, $E_n $ are the energy
eigenvalues, possibly $g_n$-fold degenerated, and $\psi_{n\alpha} (x)$
are the corresponding eigenfunctions ($\alpha=1,\ldots,g_n$).
To fix ideas we will assume the Hamiltonian to be of the
Schr\"odinger form
\be
H = - \frac{\hbar^2}{2 m} \, {\cal D}^2 + V(x)
\ee
although many of the features presented here can be generalized.
Here ${\cal D}$ represents a covariant space derivative ${\cal D}_i =
\nabla_i + \frac{i}{\hbar} A_i$ with $A_i$ the vector gauge potential
which includes the coupling constant.
The cumulative number of states $N(E)$, counting the number of states
below a certain energy $E$,  is defined as
\be
N(E)  =  \sum_{n=0}^{\infty} g_n \, \Theta(E - E_n)
 =  \Tr \Theta(E - H) ,
\label{eq.3}
\ee
where $g_n$ stands for the possible degeneracy of the system. From here
the level density can be derived to be
\be
g(E)  =  \frac{d}{dE} N(E)   =   \sum_{n=0}^\infty  g_n \, \delta(E - E_n).
\ee
Both $N(E)$ and $ g(E)$ are to be interpreted as distributions.
For our purposes it is extremely
convenient to consider the following representation of the step function
distribution
\be
\Theta(x)  =  \lim_{\epsilon \rightarrow 0^+} \, \frac{1}{\pi}\,
\Ima \log (-x+i\,\epsilon) \\
 =  \left\{
\begin{array}{lcl}
1 & \phantom{lafjl} & \mbox{if} \; x > 0\\
\frac{1}{2} & &   \mbox{if} \; x = 0\\
0 & &   \mbox{if} \; x < 0
\end{array}
\right.
\ee
the limit $\epsilon \to 0^+ $ being always understood at the end of the
calculations. The branch cut for the log function is considered to be
along the positive real axis as follows
\be
\log z = \log \left| z \right| + i \, \mathop{\mbox{arg}} z
, \qquad 0 \le \mathop{\mbox{arg}} z < 2 \, \pi\,,
\ee
which has the property
\be
\frac{1}{2}\,\log\left[ (- x + i \epsilon ) ^2 \right]
= \log ( - x + i \epsilon )
\,.
\ee
This property will be used later in the relativistic case.
With these definitions the cumulative number of states
can be written as
\be
N(E) = \frac{1}{\pi} \, \Ima \Tr \log
\left( H
- E + i \, \epsilon \right).
\ee
If we further use the identity
\be
 \Tr\log A = \log\det A
\ee
with $A$ a linear operator, the following formula is obtained
\be
N(E) = \frac{1}{\pi}\, {\rm arg}\, \det(H-E+i\,\epsilon),
\ee
i.e. the cumulative number of states is proportional to the phase of
the determinant of the operator $ H - E + i\,\epsilon $. It should be
mentioned that the definition of $N(E)$ implicitly assumes that the Hamiltonian
is bounded from below. If this is not the case one can still proceed
with the aforementioned formulas but subtracting a suitable infinite but
energy independent constant. This point will be discussed when dealing
with the Dirac equation.

\subsection{\sl Physical Quantities}

Any single sum can be obtained directly from $N(E)$. For instance, the
total number of particles below a chemical potential $\mu$ is given by
$ N( E=\mu)$. Similarly, the total energy of a system of $
N(\mu)$ fermions can be expressed as
\be
E_T(\mu)  =  \sumn g_n \, E_n \, \Theta(\mu - E_n)
 =  \int dE \, g(E) \, E\, \Theta(\mu - E).
\ee
In this particular case integration by parts yields the simpler result
\be
E_T(\mu) = \mu N(\mu) - \int dE\, N(E)\,\Theta(\mu-E),
\label{et-ex}
\ee
which is particularly interesting when evaluating the free energy.
One body local quantities, such as particle densities and energy
densities can be obtained from suitable functional derivatives with
respect to external potentials with proper quantum numbers.
In the non-relativistic case\footnote{ The results for
relativistic systems will be presented in section~\ref{sec-rel-cas}.} the
particle density of a system of $N(\mu)$ fermions is
\be
\rho(x, \mu) =
\sumn \sumal \psi_{n \alpha}^*(x) \,
\psi_{n \alpha}(x) \, \Theta(\mu -E_n).
\ee
Considering the variation of $N(E)$ in eq.~(\ref{eq.3}) under a variation
$\delta V(x)$ of the potential
\be
\delta N(E) = - \Tr\left(\delta(E-H) \delta V(x) \right)
\label{eq.16}
\ee
one readily obtains the following useful result\footnote{Note that
the result is base independent due to the sum in the eigensubspace.}
\be
\frac{\delta N(E)}{\delta V(x)} =
-  \sumn\sumal \psi_{n \alpha}^*(x) \,
\psi_{n \alpha}(x) \, \delta(E -E_n),
\ee
yielding
\be
\rho(x, \mu)  =
- \int dE \, \frac{\delta N(E)}{\delta V(x)}\, \Theta(\mu -E)\,.
\ee
The total energy density
\be
\rho_E(x, \mu) =
\sumn \sumal \psi_{n \alpha}^*(x) \,E_n\,
\psi_{n \alpha}(x) \, \Theta(\mu -E_n),
\ee
can be obtained in a similar way,
\be
\rho_E(x, \mu)  =
- \int dE \, \frac{\delta N(E)}{\delta V(x)}\, E\,\Theta(\mu -E)\,.
\ee
For the kinetic energy density (total minus potential) one has
\be
\rho_K(x, \mu) =
- \int dE \, \frac{\delta N(E)}{\delta V(x)}\,
\left(E-V(x)\right) \,
\Theta(\mu -E).
\ee
Other local densities can be also computed by introducing an external
potential with suitable quantum numbers, for instance the current density
could be evaluated by a functional derivative with respect to
the external vector potential.

\subsection{\sl Semiclassical Expansion in the Non-Relativistic
case.}
\label{sec-sem-non-rel}

As we have said any one body operator can be deduced from the
cumulative number of states which in our regularization turns out to be
proportional to the phase of the determinant of $ H-E + i\,\epsilon $. Our
aim is to perform a semiclassical expansion of those objects. Besides an
overall factor this expansion correspond to a derivative expansion in
the potential $V(x)$. Therefore in most of the expressions the
$\hbar$-dependence can be directly read off. To perform this expansion
we use the method of Chan \cite{Ch86} recently extended by two of us
\cite{Ca93} and also considered in previous work \cite{Ru93} in the
present context. In the appendix we review for completeness the main
features of the method and provide some results which will be
extensively used along this paper.

Many of the features which occur in higher order WK$\hbar $ corrections
within our formalism are best exemplified in the non-relativistic case.
Thus, let us consider the semiclassical expansion of the cumulative
number of states for a system of non-relativistic fermions with spinor
degeneracy $\nu$ interacting
with a spin independent mean field potential (the generalization
to spin dependent forces is straightforward although somewhat
more involved). This case is well known
from the study of non-relativistic atomic systems in the context of
density functional theory \cite{Pa89} and we only show it here for
pedagogical reasons.  We will do this up to
fourth order in $\hbar$ in $D$-dimensions. This particular example is
also interesting because it illustrates, in the case $D = 3$, the
meaning of apparently divergent integrals. This feature does not show up
at orders in $\hbar$ strictly lower than four.
As shown in the appendix it is convenient to consider the
function $\Delta$ given by
\be
\Delta(x,k) \equiv \frac{1}{
k^2 + 2m(V(x) - E) + i \epsilon}\,.
\ee
Furthermore we use the notation (see caption of table~\ref{t-Chan})
\bea
\Delta_i & = & \hbar \, \nabla_i \Delta \nonumber\\
F_{ij} & = & \hbar \left(\nabla_i A_j
- \nabla_j A_i\right) \nonumber \\
F^2 & \equiv & F_{ij} F_{ij} \, \qquad i, j = 1,2,\ldots,D\,,
\eea
where $F_{ij}$ represents the field strength tensor of the vector
gauge potential $A_i$ times $\hbar$.
In this particular case $\Delta $ possesses a trivial matrix structure,
which simplifies a lot the actual calculation. One has
after application of table~\ref{t-Chan} the result
\be
N(E) = N_0(E) + N_2(E) + N_4(E) + \ldots\,,
\ee
where
\bea
N_0(E) & = & \frac{1}{\pi}\,
\Ima \int \frac{d^Dxd^D k}{(2 \pi \hbar)^D}
\, \log(k^2 + 2m(V(x) - E) + i \epsilon) \nonumber \\
N_2(E) & = & \frac{1}{\pi} \, \frac{ \hbar^2}{D}\,
\Ima \int \frac{d^Dxd^Dk}{(2\pi \hbar)^D} \,
k^2 \, (\nabla \Delta)^2 \nonumber \\
N_4(E) & = & - \frac{1}{\pi} \, \frac{ 2 \hbar^4}{D(D+2)}\,
\Ima \int \frac{d^Dxd^Dk}{(2\pi \hbar)^D} \,
k^4 \, \Bigl[ - (\nabla \Delta)^4 + \Bigr.\nonumber\\
& & \Bigl. \quad \mbox{} + 2 \, (\Delta (\nabla^2 \Delta))^2
- \frac{F^2}{\hbar^4} \, \Delta^4\Bigr].
\eea
It should be noted that the $N_{2 n}(E)$ term contains the power
$\hbar^{2 n - D}$. To proceed further we make use of the integrals
\bea
\lefteqn{
\frac{1}{\pi}\,\Ima \int d^Dk \, k^s \, \log (k^2 + 2m(V(x) - E)
+ i \epsilon)  = }
\phantom{ \frac{1}{\pi} \, \Ima \int \frac{d^Dk}{(2 \pi)^D}\,
k^{2 s} \, \Delta^n }
& & \nonumber\\
& = &
\frac{2 \pi^\frac{D}{2}}{\Gamma\left(\frac{D}{2}\right)}\,
\frac{1}{s+D} \, \left(2 m(E - V)\right)^\frac{D+s}{2} \, \Theta(E -
V)\nonumber\\
 \frac{1}{\pi} \, \Ima \int \frac{d^Dk}{(2 \pi)^D}\,
k^{2 s} \, \Delta^n & = &
- \frac{1}{(4 \pi)^{\frac{D}{2}}} \,
\frac{\Gamma\left( s +\frac{D}{2}  \right)}{ \Gamma\left(
\frac{D}{2}\right)^2  \Gamma(n)} \,
(2 m)^{\frac{D}{2} + s - n} \times \nonumber \\
& &
\quad \times \frac{d^{n-s-1}}{d E^{n-s-1}}
\left[ ( E - V)^{\frac{D}{2} - 1} \,  \Theta(E - V)\right],\nonumber\\
\eea
yielding the following expression
for the different contributions to the cumulative number of states
\bea
N_0(E) & = &
\frac{\nu}{(4 \pi)^\frac{D}{2} \, \Gamma\left(\frac{D}{2}+1\right)}
\, \left(\frac{2 m}{\hbar^2}\right)^\frac{D}{2} \,
\int d^Dx \, (E - V) ^\frac{D}{2} \, \Theta(E-V)
\nonumber \\
N_2(E) & = &
- \frac{\nu}{(4 \pi)^\frac{D}{2} \, \Gamma\left(\frac{D}{2}\right)}
\, \left(\frac{2 m}{\hbar^2}\right)^{\frac{D}{2}-1} \,
\int d^Dx \, \frac{ ({\bf \nabla} V)^2}{12}
\partial_E^2 \left[(E - V)^{\frac{D}{2}-1} \, \Theta(E-V)\right]
\nonumber \\
N_4(E) & = &
\frac{\nu}{(4 \pi)^\frac{D}{2} \, \Gamma\left(\frac{D}{2}\right)}
\, \left(\frac{2 m}{\hbar^2}\right)^{\frac{D}{2}-2} \,
\int d^Dx \, \frac{1}{60} \left\{
\frac{1}{24} \, ({\bf \nabla} V)^4 \, \partial_E^5
- \frac{1}{3} \, ({\bf \nabla} V)^2 \,({\bf \nabla}^2 V)
\, \partial_E^4 +
\right.
\nonumber\\
& & \quad
\left.
\mbox{ } + \frac{1}{2} \, ({\bf \nabla}^2 V)^2 \, \partial_E^3
- \frac{5}{\hbar^4} F^2 \, \partial_E
\right\}
\left[(E - V)^{\frac{D}{2}-1} \, \Theta(E-V)\right]\,.
\nonumber\\
\eea
One can reduce the number of energy
derivatives by using the identity
\be
(\nabla_i V) \, \frac{d}{dE} f(E - V) = - \nabla_i f(E-V).
\ee
Integration by parts yields the result
\bea
N_0(E) & = &
\frac{\nu}{(4 \pi)^\frac{D}{2}\,\Gamma\left(\frac{D}{2}+1\right)}
\, \left(\frac{2 m}{\hbar^2}\right)^\frac{D}{2} \,
\int d^Dx \, (E - V) ^\frac{D}{2} \, \Theta(E-V)
\nonumber \\
N_2(E) & = &
- \frac{\nu}{(4 \pi)^\frac{D}{2} \,\Gamma\left(\frac{D}{2}\right)}
\, \left(\frac{2 m}{\hbar^2}\right)^{\frac{D}{2}-1} \,
\int d^Dx \, \frac{ ({\bf \nabla}^2 V)}{12}
\partial_E \left[(E - V)^{\frac{D}{2}-1} \, \Theta(E-V)\right]
\nonumber \\
N_4(E) & = &
\frac{\nu}{(4 \pi)^\frac{D}{2}\,\Gamma\left(\frac{D}{2}\right)}
\, \left(\frac{2 m}{\hbar^2}\right)^{\frac{D}{2}-2} \,
\int d^Dx \, \frac{1}{288} \left\{
 ({\bf \nabla}^2 V)^2 \, \partial_E^3
- ({\bf \nabla}^4 V)
\, \partial_E^2 +
\right. \nonumber\\
& &
\quad
\left.
 \mbox{ }
+ \frac{2}{5} \, (\nabla_i \nabla_j V)^2 \, \partial_E^3
- \frac{24}{\hbar^4} F^2 \, \partial_E
\right\}
\left[(E - V)^{\frac{D}{2}-1} \, \Theta(E-V)\right]\,.
\nonumber\\
\label{des-N-nr}
\eea

It is interesting to notice that these formulas produce finite results
if the derivatives with respect to the energy are evaluated after the
space integrals have been performed. If the opposite process were
considered, explicit divergencies would appear due to the singularity
caused by the classical turning hypersurface.
 After careful regularization the result would coincide with that
obtained by using the previous prescription. This point can be clearly
understood when dealing with a particular example (see next section).

According to  the general formula~(\ref{et-ex}) one can use the
expressions~(\ref{des-N-nr}) to
perform a semiclassical expansion of the total energy of $ N(\mu) $
fermions
\bea
E_T(\mu) & = & E_0(\mu) + E_2(\mu) + E_4(\mu) + \ldots
\eea
where
\bea
E_0(\mu) & = & \mu N_0(\mu)
-\frac{\nu}{(4 \pi)^\frac{D}{2}\,\Gamma\left(\frac{D}{2}+2\right)}
\, \left( \frac{2 m}{\hbar^2}\right)^{\frac{D}{2}} \,
\int d^Dx \, (\mu - V)^{\frac{D}{2}+1} \,
\Theta(\mu - V)
\nonumber\\
E_2(\mu) & = & \mu N_2(\mu) + \nonumber \\
\lefteqn{
\mbox{ }
+\frac{\nu}{(4 \pi)^\frac{D}{2}\,\Gamma\left(\frac{D}{2}\right)}
\, \left( \frac{2 m}{\hbar^2}\right)^{\frac{D}{2}-1} \,
\int d^Dx \,\frac{({\bf \nabla}^2 V)}{12} \,
(\mu - V)^{\frac{D}{2}-1} \,
\Theta(\mu - V)
} & &
\nonumber\\
E_4(\mu) & = &
\mu N_4(\mu)
-\frac{\nu}{(4 \pi)^\frac{D}{2}\,\Gamma\left(\frac{D}{2}\right)}
\, \left(\frac{2 m}{\hbar^2}\right)^{\frac{D}{2}-2} \,
\int d^Dx \, \frac{1}{288} \left\{
 ({\bf \nabla}^2 V)^2 \, \partial_\mu^2 - \right. \nonumber\\
\lefteqn{
\left.
 \mbox{ }-  ({\bf \nabla}^4 V)
\, \partial_\mu + \frac{2}{5} \, (\nabla_i \nabla_j V)^2 \, \partial_\mu^2
- \frac{24}{\hbar^4} F^2
\right\}
\left[(\mu - V) ^{\frac{D}{2}-1} \, \Theta(\mu-V)\right]\,.
} & &
\nonumber\\
\eea
Similarly to the cumulative number of states one has that $E_{2
n}(\mu)$ contains a power $\hbar^{2 n - D}$.

\subsection{\sl Application to the Non-Relativistic Harmonic
Potential}
\label{sec-non-harm}

To illustrate certain aspects of the fourth order semiclassical
expansion not present in lower orders we consider the
three-dimensional non-relativistic
isotropic oscillator defined by the central potential
\be
V(r) = {1\over 2} \, m \omega^2 r^2
\ee
and $A_i = 0 $, with eigenenergies and degeneracy given by
\be
E_n = \hbar \omega \left( n + {3\over 2 } \right) , \qquad
g_n = {1\over 2} \, ( n+1)(n+2),
\ee
respectively. Hence the cumulative number of states is given by
\be
N(E) = \nu
\sum_{n=0}^\infty {1\over 2} \, ( n+1)(n+2) \Theta ( E -E_n )
\ee
where $\nu$ represents the spin degeneracy factor.
The actual calculation of successive $\hbar $ corrections to $N(E)$
is simplified by direct application of formula~(\ref{des-N-nr}). This
yields
\bea
N_0(E) & = & \frac{\nu}{6}  \left( \frac{E}{\hbar \omega}\right)^3
\Theta(E) \nonumber \\
N_2(E) &=& - \frac{\nu}{8} \left( \frac{E}{\hbar \omega}\right) \Theta(E)
\nonumber \\
N_4(E) &=& \frac{17\nu}{1920} \, \hbar \omega \, \delta(E)\,.
\label{N-nr}
\eea
Notice that the result is finite but $N(E)$ has to be understood as a
distribution in $E$. To understand how these expressions arise let us
consider the following integral appearing in the fourth order
contribution
\bea
N_4 (E) & = & \frac{17 \nu}{240\pi}\, \hbar \omega \, \partial_E^3 \,
\int_0^\infty dx\,  x^2  \sqrt{E-x^2} \Theta ( E- x^2 ) \nonumber \\
        & = & \frac{17\nu}{240\pi}\, \hbar \omega \, \partial_E^3 \,
              \left\{ {\pi\over 16} E^2 \Theta(E) \right\} \nonumber \\
       & =& \frac{17\nu}{1920} \, \hbar \omega \, \delta(E)\,.
\eea
If we had derived with respect to the energy inside the integral,
then explicit divergencies would have occurred. After careful
regularization the sum of the divergent terms would have produced the
previous well defined result. The necessity of taking the derivative
outside the integral is not directly linked to the appearance of a
genuine distribution in the final formula. For instance, if only
two derivatives with respect to the energy were involved in the previous
calculation, the final result (proportional to $\Theta(E)$) could only
be obtained by our prescription or suitable regularization. Notice that
if the step function is ignored when deriving the integrand,
then the final result is explicitly divergent.

Similarly one can also compute the semiclassical expansion for the total
energy of a system of $N(\mu)$ fermions as defined by~(\ref{et-ex})
obtaining
\bea
E_0(\mu)& = & \frac{\nu}{8}  \left( \frac{\mu}{\hbar \omega}\right)^3
\mu \, \Theta(\mu) \nonumber \\
E_2(\mu) &=& - \frac{\nu}{16} \left( \frac{\mu}{\hbar \omega} \right)
\mu\,\Theta(\mu)
\nonumber \\
E_4(\mu) &=& - \frac{17\nu}{1920} \, \hbar \omega \, \Theta(\mu)\,.
\label{E-nr}
\eea

\subsection{\sl Reordering of the Semiclassical Expansion}
\label{sec-reo}

{}From the knowledge of the  number of fermions $N(\mu)$ below a
chemical potential $\mu$ and the total energy $E(\mu)$ one can obtain
the energy of a given number of particles A, by properly adjusting
$\mu$ so that $ A=N(\mu)$. In what follows we will discuss a
procedure within the semiclassical approach, which is advantageous for
practical calculations \cite{Kr90,Sc93}. This method can be applied both
in the non-relativistic and the relativistic cases.
Our starting point is the former semiclassical expansion for the number
of particles and the total energy
\bea
N(\mu) & = &  N_0(\mu) + N_2(\mu) + N_4(\mu)+ \ldots \nonumber \\
E(\mu) & = &  E_0(\mu) + E_2(\mu) + E_4(\mu)+\ldots\,.
\label{eq-33}
\eea
{}From the normalization condition $ A = N(\mu) $  it is clear that $\mu$
inherits a certain $\hbar $ dependence. We proceed along the lines
proposed in previous works \cite{Kr90,Sc93}, by further expanding the
chemical potential into powers of $\hbar$
\be
\mu = \mu_0 + \mu_2 + \mu_4+\ldots\,,
\label{eq-34}
\ee
with the lowest order condition
\be
N_0(\mu_0) = A\,.
\label{eq-35}
\ee
It should be clear that by doing so we are simply reordering the
semiclassical series. If the expansion is truncated at a certain
finite order in $\hbar $, the result of this reordering is consistent up
to higher $\hbar $ corrections.
One should also mention that for this expansion to be a true
$\hbar$-expansion, i.e. that $\mu_0$ is independent of $\hbar$,
one has to declare that $A$ is of order $\hbar^{-D}$ since this is the
power which appears in $N_0$. We will always assume this particular
order counting. If we insert equations (\ref{eq-34})
and (\ref{eq-35})
back into the first line of equation (\ref{eq-33})
the successive terms in the expansion
of $\mu$  are automatically fixed giving
\bea
\mu_0 & : & N_0(\mu_0) = A\nonumber \\
\mu_2 & \equiv & - \frac{N_2(\mu_0)}{N_0'(\mu_0)}
\nonumber \\
\mu_4 & \equiv & - \frac{N_4(\mu_0)}{N_0'(\mu_0)}
+ \frac{N_2(\mu_0)\,N_2'(\mu_0)}{\left[N_0'(\mu_0)\right]^2}
- \frac{N_0''(\mu_0)\,\left[N_2(\mu_0)\right]^2}{2\,\left[N_0'(\mu_0)\right]^3}
\,.
\eea
Similarly one can expand the total
energy to give
\bea
E_T(\mu) & = & {\cal E}_0(\mu_0) +
{\cal E}_2(\mu_0)  + {\cal E}_4(\mu_0)
+ {\cal O}(\hbar^{6-D})\,,
\eea
where
\bea
{\cal E}_0(\mu_0) & \equiv & E_0(\mu_0)
\nonumber \\
{\cal E}_2(\mu_0) & \equiv & E_2(\mu_0) +
\mu_2(\mu_0) \,  E_0'(\mu_0)
\nonumber \\
{\cal E}_4(\mu_0) & \equiv &  E_4(\mu_0)
+ \mu_4(\mu_0) \, E_0'(\mu_0)
+ \frac{1}{2}\,\left[ \mu_2(\mu_0)\right]^2  \,  E_0''(\mu_0)
+ \mu_2(\mu_0) \, E_2'(\mu_0)\,.
\nonumber \\
\eea
If we now apply the relationship between the number of particles and the
total energy as given by eq.~(\ref{et-ex}) one obtains after
deriving with respect to $\mu$
\bea
E_m'(\mu) & = & \mu \, N_m'(\mu)
\nonumber \\
E_m''(\mu) & = &
N_m'(\mu) + \mu \, N_m''(\mu)\,.
\eea
Using this result we get the final simple expressions
\bea
{\cal E}_0(\mu_0) & \equiv & E_0(\mu_0)
\nonumber \\
{\cal E}_2(\mu_0) & \equiv & E_2(\mu_0) - \mu_0 \, N_2(\mu_0)
\nonumber \\
{\cal E}_4(\mu_0) & \equiv & E_4(\mu_0) - \mu_0 \, N_4(\mu_0)
+ \frac{1}{2} \,  \frac{\left[N_2(\mu_0)\right]^2}{N_0'(\mu_0)}\,.
\nonumber \\
\eea
The advantages of this reordering in the total energy are
manifold. First, the chemical potential has to be adjusted only once,
namely at lowest and simplest order. Second, there is a clear order by
order separation, that allows to treat the corrections in the energy
additively. Finally, big
cancellations are avoided in practical calculations since the
cumulative number turns out to be extremely sensitive to
variations in the chemical potential.

\section{\bf Relativistic Case}
\label{sec-rel-cas}
\subsection{\sl Cumulative Number for the Dirac equation}

In this section we will consider the semiclassical expansion of the
Dirac Hamiltonian with external vector and scalar fields. The results
derived here will be applied to estimate higher WK$\hbar$ corrections
in a mean field relativistic nuclear model.

The Dirac eigenvalue problem of a particle in $D$ dimensions can be
written as
\bea
\lefteqn{
H \, \psi_{n\alpha} = E_n \, \psi_{n\alpha}
} \phantom{H} \nonumber\\
H & \equiv & \alpha \cdot P + \beta \, \Phi(x) + V(x)\,,
\eea
where $ \alpha \cdot P \equiv \sum_{i=1}^D \alpha^i P^i $ with
$\alpha^i \equiv \gamma^0 \gamma^i $
and $\beta=\gamma^0$. The gamma matrices satisfy the
usual anticommutation relation $ \{ \gamma_\mu , \gamma_\nu\} = 2
g_{\mu\nu} $ where the Minkowskian metric is $ (1, -1 , \ldots , -1 ) $
and the generalized momentum $P^i $ is given by
\be
P^i \equiv - i\, \hbar \, \partial_i - A^i(x)\,.
\ee
The time independent potentials $\Phi( x ) $ and $ V( x) $ represent
the scalar field and the time component of a (D+1)-vector field respectively.
The rest mass of the particle $m$ is included in the scalar field
through the asymptotic condition $ \Phi (\infty ) = m $. The cumulative
number of states is defined as
\be
N^H  (E)  =  \sum_n g_n \, \Theta(E-E_n)
        =  \Tr \Theta(E-H),
\ee
where $g_n$ represents the possible degeneracy of the eigenvalue $E_n$.
The Dirac Hamiltonian is not bounded from below and hence the
cumulative number becomes infinite. Nevertheless one can redefine this
quantity by suitably choosing the energy for which the number of
states is counted from. This operation corresponds to a
certain normal ordering. In the free case, the Dirac equation exhibits
an energy gap of width $ 2 m $, which distinguishes between positive and
negative energy eigenvalues. In the interacting case it seems reasonable
to assume that, if the fields $V$ and $\Phi$ are not too strong compared
to the rest mass $m$, some gap in the Dirac spectrum exists. For the
potentials which usually appear in relativistic nuclear physics this
turns out to be the case. We will
see that in the semiclassical limit there might appear an energy  gap
if the interval $ [ {\rm max} \{ V(x)-\Phi(x)\},{\rm min}\{V(x)+\Phi(x)\}]$
is not empty. The former discussion suggests the subtraction point to
lie within the gap
\be
N(E)  \equiv  N^H (E) - N^H (E_{\mbox{gap}}).
\ee
Furthermore, it is convenient to split
$ N(E)$ into the
regions above ($ \Omega_+ $) and below ($ \Omega_- $) the gap as
follows
\bea
N(E)   & = & N_+(E) - N_-(E) \nonumber\\
N_+(E) &\equiv& \sumop g_n \, \Theta(E-E_n) \nonumber\\
N_-(E) &\equiv& \sumom g_n \, \Theta(E_n-E).
\eea
It should be noticed that with this definition both $N_+ (E) $ and $ N_-
(E) $ are non-negative numbers. The level density is then given by
\bea
g(E) \equiv \frac{dN(E)}{dE} = \frac{dN^H (E)}{dE} \nonumber\\
\phantom{g(E)} = \sumo g_n \ \delta(E-E_n) \,,
\eea
where the irrelevant constant $N^H (E_{\rm gap})$ disappears.
To evaluate the semiclassical expansion of $N(E)$ we will deal
directly with $N^H(E)$ and then we will proceed to perform the
subtraction and the separation between the two branches (above and below
the gap).

\subsection{\sl Independent Particle Approximation }

Notice that all manipulations above are intrinsic to the Dirac spectrum
and do not yet define a many body picture. In the spirit of the Hartree
approximation (independent particle model) we introduce a chemical
potential as we did in the non-relativistic case. However, the total
energy computed as the sum of all eigenstates below the chemical
potential is an ultraviolet divergent quantity, which cannot be made
finite by simply subtracting the vacuum contribution from the Dirac
sea. In a renormalizable field theory, one could subtract additional
counterterms to the total energy. To keep the line of reasoning straight
we relegate the discussion of the Dirac sea corrections to section
\ref{sec-sea}. For most of this paper we will work in the Hartree
valence approximation, where many calculations have been done.
As we will show the semiclassical Dirac sea corrections can be formally
obtained from the valence contribution. From the point of view of the
many body theory this approximation corresponds to deal with a
non-relativistic many body problem using relativistic kinematics.

The valence energy of a system of $A$-fermions with energies between the
chemical potential and the energy gap is given by
\be
E_T(\mu) = \mu \, N(\mu) - \int dE\, N(E) \, \Theta(\mu-E)
\ee
with the normalization condition
\be
A = N(\mu).
\ee
In evaluating several single particle densities, it is necessary
to make use of a generalization of the eq.~(\ref{eq.16})
to the relativistic case
\bea
\frac{\delta N_\pm(E)}{\delta V(x)} & = &
\mp \sumopm \sumal \psi_{n \alpha}^\dagger(x) \,
\psi_{n \alpha}(x) \, \delta(E- E_n) \\
\frac{\delta N_\pm(E)}{\delta \Phi(x)} & = &
\mp \sumopm \sumal \psi_{n \alpha}^\dagger(x) \, \beta \,
\psi_{n \alpha}(x) \, \delta(E- E_n).
\eea
Thus, the valence energy density is given by
\bea
\rho_E(x,\mu) & = & \rho_E^+(x,\mu) - \rho_E^-(x, \mu)
\nonumber\\
\rho_E^\pm(x,\mu) & = & \sumopm \sumal \psi_{n\alpha}^\dagger(x)
\, E_n \, \, \psi_{n\alpha}(x) \, \Theta\left(\pm\left(\mu - E_n
\right)\right) \nonumber\\
&=&
\mp \int d E \, \frac{\delta N_\pm(E)}{\delta V(x)} \,
E\, \Theta\left(\pm\left(\mu - E\right) \right),
\label{eq-55}
\eea
the valence particle density is
\bea
\rho(x,\mu) & = & \rho^+(x,\mu) - \rho^-(x, \mu)
\nonumber\\
\rho^\pm(x,\mu) & = & \sumopm \sumal \psi_{n\alpha}^\dagger(x)
\, \psi_{n\alpha}(x) \, \Theta\left(\pm\left(\mu - E_n
\right)\right)\nonumber\\
&=&
\mp \int d E \, \frac{\delta N_\pm(E)}{\delta V(x)} \,
\Theta\left(\pm\left(\mu - E\right) \right),
\label{eq-56}
\eea
and the valence scalar density reads
\bea
\rho_s(x,\mu) & = & \rho_s^+(x,\mu) - \rho_s^-(x, \mu)
\nonumber\\
\rho_s^\pm(x,\mu) & = & \sumopm \sumal \psi_{n\alpha}^\dagger(x)
\, \beta \, \, \psi_{n\alpha}(x) \, \Theta\left(\pm\left(\mu - E_n
\right)\right)\nonumber\\
&=&
\mp \int d E \, \frac{\delta N_\pm(E)}{\delta \Phi(x)} \,
\Theta\left(\pm\left(\mu - E\right) \right).
\label{eq-57}
\eea
The kinetic energy density could be readily obtained from the former
ones by means of the formula
\bea
\rho_K^+(x,\mu) & = & \sumop \sumal \psi_{n\alpha}^\dagger(x)
\, (\alpha \cdot P + \beta \, m - m) \, \, \psi_{n\alpha}(x) \,
\Theta\left( \mu - E_n \right)\nonumber\\
& = &
\rho^+_E(x, \mu)  + (m -\Phi(x)) \, \rho^+_s(x,\mu)
- (m + V(x))\, \rho^+(x, \mu).
\eea

\subsection{\sl Semiclassical Expansion}

Following previous work \cite{Ru93} we can apply a semiclassical
expansion
to the relativistic case. A direct application of the logarithmic
regularization to the step function gives
\bea
N^H (E) &= &\frac{1}{\pi} \, \Ima\Tr \log(H - E + i \, \epsilon) \nonumber \\
& = & \frac{1}{\pi}\, {\rm arg} \det(H-E+i\,\epsilon).
\eea
which does not allow yet to employ the general method described in
the appendix.
To do so, we consider the transformation of the Dirac operator in an odd
number of dimensions
\bea
\tilde H & \equiv & \left(\beta \gamma_5\right) \, H \,
\left(\beta \gamma_5\right)^{-1}
\nonumber\\
 &= & - \alpha \cdot P - \beta \, \Phi(x) + V(x)\,,
\eea
where $\gamma_5 \equiv
- i \, \gamma_0 \gamma_1 \cdots \gamma_D = \gamma_5^{-1} $.
Using the properties of the determinant one obtains that
\be
\arg \det(H-E+i\,\epsilon) = \frac{1}{2}\,
\arg \det\left[(H-E+i\,\epsilon)\,(\tilde H-E+i\,\epsilon)\right],
\ee
whence
\be
N^H(E) = \frac{1}{2 \pi}\, \Ima\Tr\log(-P^2- U(X))\,,
\ee
where the following definitions have been introduced
\bea
U(X) & \equiv & \Phi^2 - (E - V- i \,\epsilon)^2 +
i \, \hbar \, \alpha \cdot \nabla(V- \beta\, \Phi) - \frac{1}{2} \,
\sigma_{ij}\, F_{ij} \nonumber \\
\sigma_{ij} & \equiv& \frac{ i}{2}\, \left[\gamma^i,\gamma^j\right]
\nonumber\\
F_{ij} & \equiv & - i \, \left[P^i,P^j\right]
=  \hbar \,(\nabla_i A^j - \nabla_j A^i).
\eea
The former equation is now ready for a semiclassical treatment, since
it involves an elliptic operator with some spinorial structure. Up to
fourth order in the WK expansion a direct application of
table \ref{t-Chan} yields
\bea
N^H (E) &=& \frac{1}{2\pi}\,\Ima \int \frac{d^D x \, d^Dk}{(2 \pi \hbar)^D}
\, \tr \left\{ \log(-k^2 - U) +
\frac{k^2 \, \hbar^2}{D}\,(\nabla \Delta)^2 - \right.\nonumber\\
& \lefteqn{
 \mbox{ }-\frac{2 \, k^4 \, \hbar^4}{D(D+2)}\,
\left[ 2 \, \left(\Delta \left(\nabla^2 \Delta\right)\right)^2
+ \left(\left(\nabla_i \Delta\right) \, \left(\nabla_j \Delta\right)
\right)^2 - 2 \, \left( \nabla \Delta\right)^4 -
\right.
} & \hfill \nonumber\\
& & \left.\left.
\mbox{ } - \frac{1}{\hbar^{4}} \, \left( F_{ij} \Delta^2\right)^2
- \frac{4}{\hbar^2}\, iF_{ij} \, \Delta (\nabla_i \Delta) \,
(\nabla_j \Delta) \, \Delta \right] + {\cal O}(P^6)\right\}\,,
\nonumber\\
\eea
where the matrix propagator $\Delta(x,k)^{-1} \equiv k^2 + U(x)$
has been defined. This operator
contains some $\hbar $ dependence, so
that we reexpand around
\be
\Delta_o(x,k)^{-1}= k^2 + \Phi^2 - (E-V-i\epsilon)^2.
\ee
Direct calculation of the Dirac traces together with integration by
parts yields
\be
N^H(E)  =  N^H_0(E) + N^H_2(E) + N^H_4(E) + \Theta(\hbar^{6-D})
\ee
where
\bea
N^H_0(E) & = & \frac{T(D)}{2\pi} \,
\Ima \int \frac{d^D x \, d^D k}{(2 \pi \hbar)^D}\, \log(-\Delta_o^{-1})
\nonumber  \\
N^H_2(E) &=&\frac{T(D)\hbar^2}{2\pi} \,
\Ima \int \frac{d^D x \, d^D k}{(2 \pi \hbar)^D}\,
\left\{
\frac{1}{2}\, \left[ (\nabla V)^2 - (\nabla \Phi)^2 \right]
\, \Delta_o^2 + \right.\nonumber\\
& & \left. \qquad \mbox{ }
+ \frac{k^2}{D} \, (\nabla \Delta_o)^2
\right\}
\nonumber  \\
N^H_4(E) &=&\frac{T(D)\hbar^4}{2\pi} \,
\Ima \int \frac{d^D x \, d^D k}{(2 \pi \hbar)^D}\,
\left\{
 - \frac{1}{4}\, \Delta_o^4 \, \left[ (\nabla V)^4 + (\nabla \Phi)^4 -
\right. \right. \nonumber\\
& & \left. \left. \quad \mbox{ }
-6 \, (\nabla V)^2  \, (\nabla \Phi)^2
+ 4 (\nabla V \cdot \nabla \Phi)^2
 \right] +
 \right. \nonumber \\
& &  \left. \quad \mbox{ }
+ \frac{k^2}{D} \,
\left[ - \left( \nabla_i \left( \Delta_o^2 \nabla_j V\right)\right)^2
       + \left( \nabla_i \left( \Delta_o^2 \nabla_j \Phi\right)\right)^2 +
\right. \right. \nonumber \\
& & \left. \left. \quad \mbox{ }
       + 2\,\Delta_o^3 \, (\nabla^2 \Delta_o) \, \left(
\left( \nabla V\right)^2 - \left( \nabla \Phi\right)^2 \right)
\right] -
\right. \nonumber \\
& & \left.\quad \mbox{ }
- \frac{2 k^4}{D(D+2)} \,
\left[
2 \, \Delta_o^2 \, (\nabla^2 \Delta_o)^2 - (\nabla \Delta_o)^4
+ \frac{2}{\hbar^4} \, F^2 \, \Delta_o^4
\right]
\right\}\,.
\nonumber \\
\eea
Here $T(D)$ is the spinor space dimension ($T(1)=2$; $T(3)= 4$).
To work out the momentum integrals we note that any power of $\Delta_o
$ can be obtained as
\be
\Delta_o^n = \frac{(-1)^{n-1}}{\Gamma(n)}\, \partial_{\Phi^2}^n
\log(- \Delta_o^{-1}).
\ee
We also use the identity
\be
\lim_{\epsilon \to 0^+} \frac{1}{\pi}\Ima \,
\log(x-i \,y \,\epsilon) = 1 + \mbox{sign}(y) \, \Theta(x),
\ee
and define the local momentum
\be
p^2(x) \equiv (E-V(x))^2 - \Phi(x)^2 \,.
\ee
It is interesting to notice that within a semiclassical treatment one
can always define a gap if the interval
$[{\rm max}\{V(x)-\Phi(x)\},{\rm min}\{V(x)+\Phi(x)\}]$
is not empty.
In practice this allows to explicitly distinguish the contribution of
states below and above the gap in the semiclassical limit by means of
the formula
\bea
\lefteqn{
1 + \mbox{sign}(E - V) \, \Theta(p^2 - k^2)   = }  \nonumber \\
& = & 1 + \Theta(E-V-\sqrt{\Phi^2 + k^2})
   - \Theta\left(-\left(E-V\right) - \sqrt{\Phi^2 + k^2}\right)
{}.
\nonumber
\eea
The momentum integrals which appear in the semiclassical expansion
have the form
\bea
I_D^{n,s}(E)  & = & \frac{1}{\pi}\,\mbox{Im}\, \int \frac{d^D k}{(2 \pi)^D}
\, k^{2 s} \Delta_o^n \nonumber\\
& = &
\frac{ (-1)^{n-1} \, 2}{(4 \pi)^{D/2}\,\Gamma(n) \Gamma\left(\frac{D}{2}
\right)} \, \partial_{\Phi^2}^n \int_0^\infty dk \,
k^{2 s + D -1} \, \times \nonumber\\
& & \quad \mbox{ } \times
 \left[ 1 + \Theta(E-V-\sqrt{\Phi^2 + k^2})
   - \Theta\left(-\left(E-V\right) - \sqrt{\Phi^2 + k^2}\right)
\right]
.\nonumber
\eea
If we further subtract the contribution from the gap and explicitly
separate the two branches we get for $n > 0 $
\be
I_D^{n,s} ( E ) -
   I_D^{n,s} ( E_{\rm gap} ) = I_D^{+,n,s}(E) - I_D^{-,n,s}(E)
\ee
where
\be
I_D^{\pm,n,s}(E) =
\frac{(-1)^{n-s} \, \Gamma\left(s + \frac{D}{2}\right)}{
(4 \pi)^{D/2} \, \Gamma(n) \, \Gamma\left(\frac{D}{2}\right)^2}
\, \partial_{\Phi^2}^{n-s-1}\left[p^{D-2} \, \Theta(\pm (E-V) - \Phi)
\right].
\ee
The zeroth order term can be evaluated following similar steps. After
suitable integration by parts, the final result for the cumulative
number of states reads
\bea
N_0^\pm (E) &= &{T(D)\over 2}{1\over (4\pi\hbar^2)^{D/2}\Gamma(\frac{D}{2}+1)}
\int d^Dx\, p^D\,\Theta(\pm(E-V)-\Phi) \nonumber\\
N_2^\pm (E) &=& -{T(D)\over 2}{\hbar^2\over (4\pi\hbar^2)^{D/2}
\Gamma(D/2)}\int d^Dx\,
\Bigl\{ {1\over 2}[(\nabla\Phi)^2-(\nabla V)^2]\partial_{\Phi^2} +
\Bigr. \nonumber\\
& &\quad \Bigl. + {1\over 3}[\Phi\,(\nabla\Phi)+(E-V)\,(\nabla
V)]^2\partial_{\Phi^2}^2\Bigr\} p^{D-2}\Theta(\pm(E-V)-\Phi) \nonumber\\
N_4^\pm (E) &=& -{T(D)\over 2}{\hbar^4\over (4\pi\hbar^2)^{D/2} \Gamma(D/2)}
\int d^D x\,
 \Bigl\{  \nonumber\\
& &{\hskip -1.5cm}\mbox{ }
\phantom{+}{1\over 90} [\Phi\,(\nabla \Phi) + (E-V)\,(\nabla V)]^4
\partial_{\Phi^2}^5
+{1\over 12}[ (\nabla^2\Phi )^2 - (\nabla^2 V)^2 ]\partial_{\Phi^2}^2 +
\nonumber\\
& &{\hskip -1.5cm}\mbox{ }
+{1\over 24} \Bigl[ (\nabla V)^4 + (\nabla \Phi)^4
-6\,(\nabla V)^2 (\nabla \Phi)^2 +4\,(\nabla V \cdot \nabla \Phi)^2 \Bigr]
\partial_{\Phi^2}^3 + \nonumber \\
& &{\hskip-1.5cm}\mbox{ }
 + {1\over 12} [(\nabla V)^2 - (\nabla \Phi)^2]
[ (\nabla \Phi)^2 + \Phi \,(\nabla^2 \Phi) - (\nabla V)^2 +
(E-V)\,(\nabla^2 V) ]\partial_{\Phi^2}^3   + \nonumber\\
& &{\hskip-1.5cm}\mbox{ }
 +{1\over 6} \left[ (\nabla^2\Phi) \, (\nabla_i \Phi)
- (\nabla^2 V) \, (\nabla_i V) \right]\,
 \left[\Phi (\nabla_i \Phi) + (E-V)(\nabla_i V) \right]
\partial_{\Phi^2}^3  +
\nonumber\\
& &{\hskip-1.5cm}\mbox{ }
+{2\over 45} \left[\Phi \, (\nabla \Phi) + (E-V) \, (\nabla V)\right]^2
\times\nonumber\\
& &{\hskip-1.5cm}\mbox{ } \quad
\times
\left[ (\nabla \Phi)^2 + \Phi \, (\nabla^2 \Phi) - (\nabla V)^2 +
(E-V) \, (\nabla^2 V) \right] \partial_{\Phi^2}^4 + \nonumber\\
& &{\hskip-1.5cm}\mbox{ }
+{1\over 30} \left[ (\nabla \Phi)^2 +\Phi \, (\nabla^2 \Phi) -(\nabla V)^2 +
(E-V)\,(\nabla^2 V) \right]^2 \partial_{\Phi^2}^3  +
\Bigr. \nonumber \\
& &\Bigl. {\hskip-1.5cm}\mbox{ }
+ {1\over 6}{F^2\over\hbar^4}
\partial_{\Phi^2} \Bigr\} p^{D-2} \Theta(\pm(E-V)-\Phi)\,,
\eea
which has been already obtained in a previous work \cite{Ru93}. In
practical
calculations it is desirable to lower the number of derivatives acting
on the step function. This can be achieved by using the identity
\bea
\lefteqn{
\nabla_i \left[ f(p)\, \Theta(\pm(E-V) - \Phi)\right]
 = } \nonumber\\
& = & 2 \left[ \Phi\, (\nabla_i \Phi) + (E-V) \, (\nabla_i V)\right]
\, \partial_{\Phi^2} \left[f(p) \, \Theta(\pm(E-V) - \Phi)\right]
,
\nonumber\\
\eea
together with integration by parts yielding
\bea
N_0^\pm (E) &= &{T(D)\over 2}{1\over (4\pi\hbar^2)^{D/2}\Gamma(\frac{D}{2}+1)}
\int d^Dx\, p^D\Theta(\pm(E-V)-\Phi) \nonumber\\
N_2^\pm (E) &=& {T(D)\over 2}{\hbar^2\over
(4\pi\hbar^2)^{D/2}\Gamma(D/2)}\int d^Dx\,
\Bigl\{ {1\over 2}[(\nabla V)^2-(\nabla \Phi)^2] -
\Bigr. \nonumber\\
& &\quad \Bigl. - {1\over 12} \nabla^2 p^2
\Bigr\} \, \partial_{\Phi^2} p^{D-2}\Theta(\pm(E-V)-\Phi) \nonumber\\
N_4^\pm (E) &=& {T(D)\over 2}{\hbar^4\over (4\pi\hbar^2)^{D/2} \Gamma(D/2)}
\int d^D x\,
 \Bigl\{  \nonumber\\
& &  {\hskip-1.5cm} \mbox{ }
 - {1\over 288}\,[ (\nabla^2 p^2)+ 12 \, (\nabla \Phi )^2 - 12\, (\nabla V)^2 ]
\, (\nabla^2 p^2) \, \partial_{\Phi^2}^3
- {1 \over 720}\, (\nabla_i \nabla_j p^2)^2 \partial_{\Phi^2}^3
-
\nonumber\\
& &  {\hskip-1.5cm} \mbox{ }
- {1\over 24}\, \Bigl[ (\nabla V)^4 + (\nabla \Phi)^4
-6\,(\nabla V)^2 \, (\nabla \Phi)^2 +4\,(\nabla V\cdot\nabla \Phi)^2 \Bigr]
\partial_{\Phi^2}^3 + \nonumber \\
& & {\hskip-1.5cm}
\mbox{ }+{1\over 288} \, (\nabla^4 p^2) \,\partial_{\Phi^2}^2
+ {1\over 12}
[   (\nabla_i\Phi)\,(\nabla_i \nabla^2 \Phi)
  - (\nabla_i V)\,(\nabla_i \nabla^2 V) ]\partial_{\Phi^2}^2   - \nonumber\\
& & {\hskip-1.5cm}
\Bigl.
\mbox{ } -{1\over 6}{F^2\over\hbar^4}
\partial_{\Phi^2} \Bigr\} p^{D-2} \Theta(\pm(E-V)-\Phi)\nonumber\\
\eea
Again, the final result for $N(E)$ is explicitly convergent if the
derivatives with respect to the scalar field are evaluated after space
integration, as can be seen by introducing a new parameter $\lambda $
\bea
\lefteqn{
\int d^D x \, g[x,\Phi] \, \partial_{\Phi^2}^n \left[
f[\Phi^2]\, \Theta(\pm(E-V)-\Phi) \right] = } \nonumber\\
& =
\left. \partial_{\lambda^2}^n \int d^D x \, g[x,\Phi]\, f[\Phi^2 + \lambda^2]
\, \Theta(\pm(E-V)-\sqrt{\Phi^2 + \lambda^2}) \right\vert_{\lambda=0}
\,.
\eea
It is interesting to notice that the non-relativistic limit of the
former expressions yields that one given in
section~(\ref{sec-sem-non-rel}).
The only difference stems from the spin interaction with the gauge field.
For instance for $D=3$ the correct non-relativistic limit of the Dirac
equation is not the Schr\"odinger equation but the Pauli equation which
contains a dipole spin interaction $- \frac{\hbar}{2 m} \,
{\bf \sigma}\cdot (\nabla \wedge {\bf A})$. This has its
corresponding effect in the magnetic polarizability terms, since the
pure non-relativistic treatment gives
\be
N_4^{\rm mag}(E) = - \frac{1}{48 \, \pi^2 \, \hbar^3 \, \sqrt{2 m}}\,
\int d^3 x \, \frac{F^2}{(E-V)^{1/2}}\, \Theta(E - V)\,,
\ee
and the
non-relativistic limit of the Dirac equation produces
\be
N_4^{\rm mag}(E) =  \frac{1}{24 \, \pi^2 \, \hbar^3 \, \sqrt{2 m}}\,
\int d^3 x \, \frac{F^2}{(E-V)^{1/2}}\, \Theta(E - V)\,.
\ee

The explicit expressions for the upper branch contribution to the
various densities are rather lengthy and are given in tables 1, 2,
and 3, where the definitions
\bea
\epsilon_F(x, \mu) & \equiv & \mu - V(x) \nonumber \\
k_F(x, \mu) & \equiv & \sqrt{\epsilon_F^2 - \Phi(x)^2}
\nonumber \\
x_F(x, \mu) & \equiv & \frac{\epsilon_F}{k_F} \nonumber \\
l_F(x, \mu) & \equiv & \log{\frac{\epsilon_F + k_F}{\Phi(x)}}\,.
\eea
have been used.
A few remarks are in order. The expressions have been written in
the simplest possible way and following the standard fashion. However,
they appear to be explicitly divergent at the classical turning
hypersurface and they would produce non convergent mean values.
In fact, when writing those expressions, some purely distributive
terms have been intentionally omitted. They can be recovered by means of
the rules
\bea
x_F & \to & \frac{\mu-V}{k_F}\nonumber\\
k_F^{-n} \, \Theta(\mu - V - \Phi)
& \to & -2^n \,
\frac{\Gamma\left( \frac{n+1}{2}\right)}{\Gamma(n)}\,
\partial^{\frac{n+1}{2}}_{\Phi^2} \left[
k_F \, \Theta(\mu - V - \Phi)\right]
\label{sub}
\eea
(notice that $n$ is always an odd number so that the derivative is
evaluated an integer number of times).
It can be checked that then any negative power of $k_F $ produces finite
results for the mean values of the densities although the density
itself still remains divergent at the classical turning
hypersurface. Similarly
to the cumulative number it is convenient to replace the scalar field
derivatives by introducing an external parameter $\lambda $ so that
the space integration can be performed first. Obviously such a procedure
becomes rather involved in practical numerical applications. In a
forthcoming section
we will present a trick specially suited for numerical calculations and
valid in the case of central potentials, that avoids the use of the
substitution rules given above, hence allowing for a literal usage of
tables \ref{t-den-bar}-\ref{t-den-en}.

Finally, let us remark that the case $\Phi=m=\rm{constant}$, $A=0$,
corresponds to a relativistic atomic system. For such a system
the semiclassical
expansion up to $\hbar^4$ order has been performed \cite{Po91} in the
spirit of density functional theory. For this particular case
our formulas in table~\ref{t-den-bar} and~\ref{t-den-en} coincide
(up to purely distributional terms from eq.~(\ref{sub})) with
their intermediate WK results.

\subsection{\sl Application to Relativistic Harmonic Potentials }

In previous work \cite{Ce90,Ce93b} harmonic potentials in the Dirac
equation
have been used to illustrate the WK expansion up to second
order in $\hbar$ in the total energy, and a comparison with the exact
results as well as with the Strutinsky average \cite{St67,Br73} have
been carried
out. At this order of $\hbar$ many of the problems discussed above
do not take place, in particular the distributional character of the
cumulative number. In this section we proceed to apply our
formalism to the harmonic potentials up to fourth order in $\hbar$
and for $D=3$ space dimensions along the lines of \cite{Ru93}.
More specifically we consider the central potentials
\be
V(r) = \Phi(r) - m = \frac{1}{4} \, m\omega^2 r^2\,, \qquad A_i(r) =
0\,.
\ee
The Dirac eigenvalue problem can be solved for the bound states of
positive energy. The energy of the n-th excited state reads
\be
(E_n^+-m)({E_n^+}^2 - m^2)=
2 m \omega^2 \, \left(n+\frac{3}{2}\right)^2\,,
\ee
and has degeneracy $ g_n = (n+1)(n+2) $, hence the cumulative number of
states reads
\be
N^+(E) = \sumn (n+1)(n+2)\, \Theta(E-E_n^+)\,.
\ee
Straightforward application of our formulas yields
\bea
N_0^+(\mu) & = & \frac{8}{3}\, m^3 \,
\left(\frac{m}{\hbar \omega }\right)^{3} \, (1 + x)^{3 \over 2}
\, x^3 \, \Theta(\mu - m) \nonumber \\
N_2^+(\mu) & = & - \frac{1}{2}\, m \,
\left(\frac{m}{\hbar \omega }\right) \, (1 + x)^{1 \over 2}
\, x \, \Theta(\mu - m) \nonumber \\
N_4^+(\mu) & = & \frac{17}{960}\, \hbar \omega \,
\delta(\mu - m) \,,\nonumber \\
\eea
where the scaled dimensionless variable $x= (\mu-m)/(2m)$ has been introduced.
Note that the fourth order term is a pure distribution. Similarly the
total energy up to fourth order reads
\bea
E^+_0(\mu) & = &
\frac{512}{3465}\, m \,
\left(\frac{m}{\hbar \omega}\right)^{3} \, (1 + x)^{1\over 2}
\, \left[\frac{945}{32}\, x^5 +\frac{2905}{64}\, x^4 +
\right. \nonumber\\
& & \left. \quad \mbox{ }
+\frac{1135}{64}\, x^3 +\frac{3}{8}\, x^2 -
\frac{1}{2}\, x +1 - (1+x)^{-{1 \over 2}}
\right] \, \Theta(\mu-m)\nonumber\\
E^+_2(\mu) & = &
- \frac{4}{15}\, m \,
\left(\frac{m}{\hbar \omega}\right)\, (1 + x)^{1\over 2}
\, \left[\frac{9}{4}\, x^2 +\frac{11}{8}\, x +1 -
\right. \nonumber\\
& & \left. \qquad \mbox{ }
- (1+x)^{-{1 \over 2}}
\right] \, \Theta(\mu-m)\nonumber\\
E^+_4(\mu) & = &
-\frac{17}{960}\, \hbar \omega
\, \left[ \Theta(\mu-m) - m \, \delta(\mu - m)\right]\,.
\eea
It is interesting to note that the distributive contribution in the
fourth order term corresponds to the fourth order correction of $ m
N(\mu)$ which is the total energy of $N(\mu)$ particles and hence
cancels in the expression for the binding energy. We ignore whether this
is a general feature of the semiclassical expansion. As a side remark
let us mention that the non-relativistic reduction of the former
expressions coincides with the non-relativistic result of section
\ref{sec-non-harm}, corresponding to the spin degeneracy $\nu=2 $ and
a non-relativistic potential $ V+ \Phi -m $, provided the rest mass
contributions for the energy and the chemical potential are subtracted.

The semiclassical expansion for this particular example can be checked
directly (as suggested in \cite{Kr90,Br73}) by means of the asymptotic
Euler-Maclaurin summation formula \cite{Ab70}
\be
\sumn F(n) = \int_0^\infty dt \, F(t) + \frac{1}{2} \, \left[F(0) +
F(\infty)\right] + \sum_{n=1}^{\infty} \frac{B_{2n}}{(2 n)!}\,
\left[ F^{(2n-1)}(\infty) - F^{(2n-1)}(0)\right]\,,
\ee
where $B_m $ stand for the Bernoulli numbers.

To have a quantitative idea of how large are the corrections we take
$m =  939$ MeV and fix the oscillator constant through the relation
$\hbar \, \omega = 41 A^{-{1 \over 3}}$ MeV.
We also consider two nucleon species so that $A=N+Z$ and $N=Z$. Our
results
are presented in table~\ref{t-arm}. We consider two versions of the
semiclassical
expansion and compare them to the exact result and the Strutinsky
average. In any case we consider the total number of particles to be
$A=N(\mu)$. In the first version of the semiclassical expansion, denoted
as non-perturbative (NP), we proceed as follows. For a given order in
$\hbar$, say $\hbar^{2k}$ we fix numerically the chemical potential
$\mu$ directly from
the normalization equation $ A= N_0(\mu) + \ldots +N_{2k}(\mu) $
and recompute the
energy up to that order as $E(\mu) = E_0(\mu) + \ldots + E_{2k}(\mu)$.
This is done
independently for each order. In the second method, denoted as
perturbative (P), we proceed along the lines described in
section~(\ref{sec-reo}) \cite{Kr90,Sc93}.
We fix the zeroth order condition $ A = N_0 (\mu_0 ) $ and compute the
corrections in $\hbar$ to it by expanding $\mu = \mu_0 + \mu_2 + \ldots $. This
second procedure has the practical advantage of making the
contributions of different orders to be additive. We also construct the
Strutinsky averaged (SA) energy since it is commonly believed that the
semiclassical expansion should converge to it, and not to the exact
energy since shell effects have been smoothed out. To compute
this average we have
chosen a Gaussian weight function. As can be seen from the table
the semiclassical expansion seems indeed to converge to the Strutinsky
average
 in this particular example
at least up to order WK$\hbar^4 $. We also observe a clear
trend
between the non-perturbative and the perturbative determination of the
chemical potential, namely there appears not to be a substantial
improvement in the numerical value of the energy. However, one has to
say that the perturbative method avoids big cancellations as it turns
out to be the case in the non-perturbative procedure, and hence it is
better suited for numerical estimates.
Incidentally, let us mention that our non-perturbative calculation up to
WK$\hbar^2$ coincides numerically with previous estimates \cite{Ce93b}
(we only find a small deviation from our Strutinsky average\footnote{We
stress the fact that the number of significant digits
presented in our table are compatible with the so called plateau
condition \cite{Br73}.}
to theirs in the case $A=16$).
Finally, let us point out that the fourth order contribution is
much smaller
($\sim 0.5 $ MeV) than
the difference between the semiclassical expansion and the exact result
($ \sim 10-20 $ MeV). This is however not the case for the second order
in $\hbar$ which turns out to be even four times larger than the shell
effects.

\subsection{ \sl Semiclassical Dirac Sea Corrections}
\label{sec-sea}
As already mentioned, the Dirac sea contribution to the energy, defined
as the sum of all Dirac eigenvalues below zero energy, is an
ultraviolet divergent quantity which cannot be made finite by simple
subtraction of the vacuum energy. However, the finite contributions of
the sea energy acquire a precise meaning within a given renormalization
scheme in a renormalizable quantum field theory, the infinite parts
corresponding to
absorb the infinities in the parameters of the Lagrangian. This is the
case for instance in the $\sigma$-$\omega$ model. Nevertheless, it has
been shown that for relativistic Lagrangians of fermions interacting
through Yukawa like couplings, the renormalization program cannot be
carried out consistently within the approximation scheme developed so
far. Indeed, a short distance vacuum instability of Landau type takes
place at the zero boson loop one fermion loop level \cite{Pe86,Co87},
so that tachyonic poles appear. It is not clear whether
this is a true feature of this kind of Lagrangians or simply an artifact
of the approximate method used to derive it. This seems to suggest
that it is more sensible to stick within a valence approximation. On
the other hand, such an approximation breaks unitarity and Lorentz
covariance (see the discussion in \cite{Se92} and references therein).
Since we are dealing with an effective theory it is not
clear what is better, to neglect the Dirac sea or to face vacuum
instabilities. In this section we consider for future reference the
WK$\hbar^4$-corrections to the Dirac sea densities.

{}From the point of view of the semiclassical expansion, such a
short distance instability can never be seen, due to the large distance
nature of the semiclassical approach. In addition, the ultraviolet
divergencies correspond to well defined orders in $\hbar$, namely the
orders WK$\hbar^0$ and WK$\hbar^2$ present quadratic and logarithmic
divergencies respectively, and higher orders turn out to be convergent.
In what follows we will derive the WK$\hbar^4$ corrections to the sea
energy from the explicit expressions obtained within the valence
approximation. First, we note that charge conjugation symmetry implies
\be
N_-(E,V)=N_+(-E,-V)
\ee
Using this equation and (\ref{eq-55}-\ref{eq-57}) the relations for
the baryon, scalar and total energy densities are
\bea
\rho^+ (x,\mu,V) &=& \rho^- (x,-\mu,-V) \nonumber\\
\rho^+_s (x,\mu,V) &=& -\rho^-_s (x,-\mu,-V) \nonumber\\
\rho^+_E (x,\mu,V) &=& -\rho^-_E (x,-\mu,-V)
\eea
respectively. In addition, the Dirac sea baryon density is given by
\be
\rho^{\rm sea} (x) = \sum_{n\in\Omega} \sum_{\alpha=1}^{g_n}
\psi_{n\alpha}^\dagger (x) \psi_{n\alpha}(x) \Theta(-E_n)
\ee
Similar expressions hold for the scalar and total energy densities.
Under the assumption that the zero energy belongs to the gap
we can effectively drop the step function and restrict the summation to
the lower branch $\Omega_-$.
This corresponds to take the limit $\mu \to
-\infty $ in the lower branch densities.
In this limit each Dirac sea density has a well defined
parity under the charge conjugation operation $V \to -V$. The
former discussion amounts to the following final equations
\bea
\rho^{\rm sea} (x) &=& - \rho^+ (x, \mu=+\infty) \nonumber\\
\rho^{\rm sea}_s (x) &=& - \rho^+_s (x, \mu=+\infty) \nonumber\\
\rho^{\rm sea}_E (x) &=& - \rho^+_E (x, \mu=+\infty)
\eea
Direct use of tables  \ref{t-den-bar}, \ref{t-den-is} and \ref{t-den-en}
yields our final result for the WK$\hbar^4$ contributions to the Dirac sea
densities. The limit $\mu\to\infty$ corresponds to $k_F\to\infty$ and
$x_F\to 1$. The densities can be looked up at table \ref{t-den-sea} for
the case in which no external gauge potential is present.
We have checked that the $\hbar^4$ Dirac sea corrections to the
baryon density can be written as a total divergence. As a consequence
the baryon number remains unchanged.
In the particular case
of constant scalar field $\Phi = m $ our formulas correspond to the sea
corrections of a relativistic atomic system already considered in ref.
\cite{Po91}. We have checked that all our results agree with theirs in
this particular case.\footnote{ Notice that our definition of
$\rho_E-V\rho$  corresponds to the density $\tau$ in that reference.}
In addition, we provide the scalar density.

\section{\bf Relativistic Nuclear Potentials }

\subsection{ \sl The $\sigma$-$\omega$ Model }

In this section we analyze the semiclassical expansion in the framework
of a mean field approximation to the $\sigma$-$\omega $ relativistic
Lagrangian \cite{Wa74} up to fourth order in $\hbar$.
We would like to stress that this model is extremely simple as a realistic
description of finite nuclei \cite{Se86,Re89,Se92}. However, it adjusts
some global properties
and reproduces the correct shell structure. Hence we feel it is a
convenient starting point to estimate the accuracy of the semiclassical
method. To be more specific, we consider the Lagrangian density to be
given by (we consider $\hbar=c=1$)
\bea
{\cal L} & = & \overline \Psi \left[ \gamma_\mu ( i \partial^\mu - g_v V^\mu)
- (M - g_s \phi) \right] \Psi + \frac{1}{2}\, (\partial_\mu \phi
\partial^\mu \phi - m_s^2 \, \phi^2) + \nonumber \\
& & \quad \mbox{ }- \frac{1}{4} \, F_{\mu \nu} F^{\mu \nu}
+ \frac{1}{2} \, m_v^2 V_\mu V^\mu\,,
\eea
where
\be
\qquad F_{\mu \nu} \equiv \partial_\mu V_\nu -\partial_\nu V_\mu \,.
\ee

The field $\Psi$ represents the nucleon field  $\Psi={\psi_P \choose
\psi_N}$,
$\phi$ stands for the scalar field and $V_\mu$ is the vector field.
$g_v$ and $g_s$ are dimensionless coupling constants. This Lagrangian
may be supplemented with counterterms since it is a renormalizable one.
We will always work in the Hartree valence
approximation (i.e. no sea approximation), where the scalar and vector
fields are considered
to be classical and the nucleus wave function is made out of a
Slater determinant of $A$ single particle states.
We refrain in the present work from making numerical estimates of the
Dirac sea corrections. Actually there are no Hartree calculations for
finite nuclei which fully include the polarization of the Dirac sea to
compare with. In addition one has the vacuum instability problems
mentioned in the previous section.
The time independent equations defining this approximation in the case
of spherical nuclei read
\bea
&
(\nabla^2 - m_s^2) \phi({\bf r})  =   - g_s \rho_s({\bf r})
\equiv -g_s \average{\Psic \Psi}_{{\rm val}} \nonumber\\
&
(\nabla^2 - m_v^2) V_0({\bf r})  = - g_v \rho({\bf r})
\equiv -g_v \average{\Psic\gamma_0\Psi}_{{\rm val}} \nonumber \\
&
\left[ -i {\bf \alpha} \cdot \nabla + g_v V_0 + \beta (M- g_s \phi)
\right] \psi_{n \pm}({\bf r}) = E_{n \pm} \, \psi_{n \pm}({\bf r})
\nonumber \\
&
V_i({\bf r}) = 0,\quad i =1,2,3\,.
\eea

Following the standard procedure we solve this set of equations
self consistently until convergence is achieved.

Finally the total energy of the system which corresponds to the
total nuclear mass can be decomposed as a sum of a fermionic
and a bosonic part, as follows
\bea
E & = & E_F + E_B \nonumber \\
E_F&=& \sum_{\alpha \in \Omega_+} (2 j_\alpha + 1) \, E_\alpha
\nonumber\\
E_B & =& \frac{1}{2} \, \int d^3x \, \left[ (\nabla \phi)^2 + m_s
\phi^2 \right] - \frac{1}{2} \, \int d^3x \, \left[ (\nabla V_0)^2 + m_v
V_0^2 \right].
\eea
The semiclassical expansion will be applied to the fermionic
part of the total energy.

\subsection{\sl Fixing of the Parameters}

We adjust the parameters of the model in the Hartree valence
approximation to nuclear matter properties following \cite{Ho81b,Se92}.
This gives
\be
C^2_s \equiv g_s^2 \, \frac{M^2}{m_s^2} = 357.4\,, \qquad
C^2_v \equiv g_v^2 \, \frac{M^2}{m_v^2} = 273.8\,,
\ee
which corresponds to a Fermi momentum $ k_F= 1.30\,\mbox{fm}^{-1}$
and a binding energy per nucleon
$B/A=15.75 \, \mbox{MeV}$. We also fix the
nucleon and vector meson masses to their physical values
$M = 939$ MeV, $m_v\equiv m_\omega = 783$ MeV.
This leaves only one free parameter $m_s $ which we adjust to
reproduce the charge mean squared radius of ${}^{40}$Ca (3.482 fm).
In summary, the set
of parameters to be considered is
\be
\begin{array}{ccc}
M = 939 {\rm MeV}, & m_s = 449.7 {\rm MeV}, & g_s = \phantom{0}9.0547,
\\[2ex]
              & m_v = 783\phantom{.0}  {\rm MeV}, & g_v = 13.7997.
\end{array}
\ee
In what follows we will apply only this parameter set. We emphasize
once more that we do not pretend to make a fully realistic description
of nuclear properties, but rather to give an idea of how large are
the WK$\hbar^4$ corrections to a pure Relativistic Fermi Gas Model.

To have a quantitative idea of their
magnitude we might plug the self-consistent scalar and vector potentials
into the semiclassical formulas obtained in
section~(\ref{sec-rel-cas}).
As it has been
already mentioned, it is necessary to compute the space integrals after
an external parameter has been introduced. For numerical computations
this is a rather unstable procedure, since in some cases one has to
evaluate up to a third derivative with respect to the external
parameter. In the next section we will propose a method to circumvent
this problem, at the expense of extending the radial dependence of the
potentials to a suitable contour in the complex plane encircling the
turning point. Strictly speaking
one might do so by solving the Hartree mean field equations along this
path. In our view this is an extremely complicated task, given the fact
that we only want to make a numerical estimate of the corrections.
We propose instead an alternative procedure based in adjusting the
self consistent potentials to a given analytical expression which can
be then trivially extended to the complex plane. In practice we propose
Woods-Saxon potentials of the form
\be
\tilde U = \frac{U_o}{1 + \exp \left\{\frac{r - R_o}{a_o}\right\}}
\label{WS-pot}
\ee
to adjust the self consistent potentials $U_V(r) = g_v V(r)$ and
$U_\phi(r) = g_s \phi(r)$.  To do so we make a least
square fit minimizing
\be
\sigma = \sqrt{ \frac{\sum_{i=1}^N \left( \tilde U(r_i) - U(r_i)
\right)^2}{N}}\,.
\ee
The result of such a fit for N = 49 equidistant points
between the origin and $R=12$ fm is given in
table~\ref{t-fit}.
To have a more quantitative idea one can look at some nuclear properties
such as binding energies\footnote{In no case have we extracted the
center of mass motion.} and radii as presented
in table~\ref{t-prop} \cite{Ja74,Va72,Wa88}. As we see the fit is
satisfactory for the computed properties.

\subsection{\sl Extension of the Potentials to the Complex Plane}

For the calculation of the total energy we  always use the perturbative
scheme, namely we adjust the chemical potential at zeroth order and
expand around this point as explained in section~\ref{sec-reo}. In
general since we deal with two
nucleon species we consider a proton and a neutron chemical potential.
Furthermore, we also consider the radial dependence of the potentials.
The basic integral appearing in the resulting expressions reads
\be
I(\lambda) =
\int dr \, g(\Phi,V,r) \, \sqrt{(\mu - V(r))^2 - \Phi(r)^2 - \lambda^2}
\, \Theta\left(\mu - V(r) - \sqrt{\Phi(r)^2 + \lambda^2}\right)
\,
{}.
\label{eq-int-r}
\ee
Other integrals can be obtained from it through successive derivatives
with respect to the external parameter $\lambda^2$ at the
particular value $\lambda=0$. The potentials we are dealing with,
present a turning point at a finite value of the radial coordinate
 $r_c(\lambda)$, given by the equation
\be
\mu - V(r_c(\lambda)) = \sqrt{\Phi(r_c(\lambda))^2 + \lambda^2}\,.
\ee
We assume, as it is often the case, that the function $g$ is
analytical in a region of the complex plane which includes the
integration interval $[0,r_c (\lambda) ] $. We also choose the branch
cut of the square root function along the real positive axis
\be
\sqrt{z} = \left\vert z \right\vert^\frac{1}{2} \,
  e^{i \left( \frac{1}{2}{\rm arg} z + \pi \right)} \,,
\qquad {\rm arg}\, z \, \epsilon
[0, 2 \pi[\,.
\ee
With this choice the integral~(\ref{eq-int-r}) can be transformed into an
integral along the path defined by the lines $ C_+ \cup C_- $ depicted
in figure~\ref{f-complex}. Under the assumption that no further
singularities occur within the closed contour
$C_+ \cup C_2 \cup C_- \cup C_1$
it is clear that the original integral
can be evaluated along the external line $ C_1 $ of the contour, i.e.
\bea
I(\lambda) & = & \frac{1}{2} \, \int_{C_+ \cup C_-} dz \,
g(\Phi,V,z) \, \sqrt{(\mu - V(z))^2 - \Phi(z)^2 - \lambda^2}
\nonumber\\
  & = & - \frac{1}{2} \, \int_{C_1} dz \,
g(\Phi,V,z) \, \sqrt{(\mu - V(z))^2 - \Phi(z)^2 - \lambda^2}\,.
\label{eq-int-C}
\eea
In this representation, the turning points never lie on the integration
path, and hence the integration and the derivatives with respect to the
external parameter $\lambda $ at $\lambda =0 $ can be interchanged. In
this case no explicit distributive contribution appears, provided we
keep away from the turning points. It should be mentioned that for
quantities which involve logarithms there might appear in principle
a different analytical structure. A detailed analysis reveals that also
in this case formula~(\ref{eq-int-C}) remains valid.
The discussion above, allows us
to readily use the results of tables~\ref{t-den-bar}-\ref{t-den-en} in
the radial case literally
with the additional modification of removing the step function and
understanding the radial coordinate to lie along path $C_1$ in the
complex plane. Finally, it is important to check that no additional
singularities appear when the path $C_1 $ is deformed. We do this in
practice by comparing our prescription (\ref{eq-int-C})
with the integral calculated
along the real axis, when no derivative with respect to the external
parameter appears. From the point of view of the complex plane, this
condition is sufficient since derivatives in the external parameter do
not introduce new singular points.

\subsection{\sl Numerical Results }

We consider the Woods-Saxon potentials described above and apply the
contour prescription explained in the previous section. The results are
presented in table~\ref{t-WS-res}. There we show the numerical values
for the total fermionic energy
at any given order, together with the exact results (WS) for those
potentials as well as the Strutinsky averaged energy (SA). The latter
is computed in the cases where it can be easily extracted, i.e. when
there are enough bound states over the chemical potential. In no case
have we considered center of
mass corrections. In general we
observe similar trends as in the harmonic potentials case, with the
exception that for lighter nuclei the convergence of the semiclassical
series seems to be slower. It is not clear whether this
behaviour corresponds to that of an asymptotic series. Finally, it is worth
mentioning that in the cases where the Strutinsky average is well
defined, the semiclassical expansion seems to converge to a value
compatible with the uncertainty inherent in the Strutinsky average.
The conclusions drawn for the harmonic potentials seem to hold also in
the case of Woods-Saxon potentials; whereas the WK$\hbar^2$ corrections to
the Fermi Gas Model provide a substantial improvement to the total
energy, this is no the case for the WK$\hbar^4$ corrections, which turn
out to be much smaller than pure shell effects.

\section{\bf Conclusions and Perspectives}

In the present paper we have studied the role of $\hbar^4 $ corrections
to the extended WK approximation within the relativistic
approach to nuclear physics. For this purpose, we have considered the
cumulative number and level density of the Dirac operator with scalar
and vector potentials as the basic objects to which the semiclassical
approximation can be applied. This allows to compute any one-body
operator within a semiclassical treatment and more important, reduces
the calculational effort considerably. We have found extremely
convenient to relate the cumulative number of states to the phase of
the determinant of the Dirac operator. Furthermore, we transform the
whole Dirac problem into a Schr\"odinger type one with some additional
spinor structure, which makes the even
nature of the $\hbar$ expansion more evident.  In fact, our
methodology reproduces previous $\hbar^2 $ calculations \cite{Ce90,We91}
rather quickly.
We have then obtained explicit  expressions for the total energy,
scalar and particle densities up to
$\hbar^4 $-order. We have found that, for central potentials, a precise
meaning to the densities can be given by considering them along a path
in the complex radial coordinate which starts and finishes at $r=0$ and
encloses the classical turning points. The mean values of such densities
produce convergent results, at the expense of extending the scalar and
vector potentials to the complex r-plane. We have also shown how the
WK$\hbar^4$ corrections for the Dirac sea can be deduced from the valence
densities and their explicit form has been given.

To have a quantitative estimate of the WK$\hbar^4 $ corrections, we have
considered the relativistic $\sigma$-$\omega$ Lagrangian in the Hartree
valence approximation. We have applied a semiclassical
expansion to the fermionic part of the total nuclear mass and
determined the contribution  of the different orders in $\hbar$. As
external profiles we have considered a Woods-Saxon fit to the exact
self-consistent Hartree solution, allowing the aforementioned extension
to the complex plane. We have compared the semiclassical expansion with
the exact result and also with the corresponding Strutinsky average in
the cases where the latter can be easily extracted. We have found that
in those cases the Strutinsky average produces, within its inherent
uncertainty, quite accurately the semiclassical WK results up to fourth
order. In general
we have found that pure shell effects are far more important than the
$\hbar^4 $-corrections, but of comparable size as the $\hbar^2 $
contributions. Nevertheless, one should say that our statement is
strictly true for the total energy which makes out the nuclear mass,
but does not necessarily imply that higher $\hbar$ corrections are
completely irrelevant for other observables, like e.g. form factors at
not too low momentum transfers.
It should be also mentioned that our results do not imply either
that the $\hbar^4$ order correction to the extended TF approximation
has to be small. In fact, as found in ref.~\cite{Ce90b} in a non-relativistic
framework, the WK expansion seems to converge faster than a density
functional approach.

Although our method has been applied to the $\sigma$-$\omega$
Lagrangian,
it can be directly generalized to more realistic and sophisticated
relativistic Lagrangians including Coulomb interaction, $\rho$ exchange,
etc. which correctly reproduce a wealth of nuclear data. We do
not expect our results to be substantially modified by such
generalization.

Our numerical calculation requires further approximations besides a pure
WK expansion. The need to produce convergent results requires
an analytical continuation of the potentials to the complex radial
coordinate plane. Given the fact that the self-consistent potentials are
available on a numerical grid, we have chosen to fit analytical
Woods-Saxon expressions to them, allowing a direct extension to the
complex plane. One could perform better, by solving the Hartree
equations on a given path of the complex r-plane.

In no case have we considered a direct minimization of the semiclassical
energy as a functional of the bosonic fields. Besides our wish to keep
the problem manageable, we do not think that this is the correct way of
proceeding, specially when the
$\hbar^4 $ corrections are included. These corrections include fourth
order derivatives which very strongly influence the behaviour of the
system at short distances, require additional boundary conditions not
present in the original formulation of the problem and might even
spoil the mere existence of solutions. At this point there is a dramatic
difference with the $\hbar^2 $ corrections. In this case the number
and form of the derivatives coincide with those of the pure classical
bosonic energy. A systematic treatment of this question for higher order
corrections might be achieved following the suggestion of
ref.~\cite{Sc93,Kr90} where
the solutions are expanded in powers of $\hbar$ around the Thomas-Fermi
solution, combined with our complex plane prescription. A detailed study
of this problem is left for future research.

In summary, we might conclude by saying that our results suggest a clear
and compelling calculational scheme in the semiclassical approach to
relativistic nuclear physics. The WK$\hbar^0$ approximation
already explains many of the gross features and substantial improvement
can be achieved by computing the next to leading order corrections,
i.e. $\hbar^2$ in the WK expansion.
{}From there on, the semiclassical expansion is not only difficult to
handle, but also accounts for a tiny fraction of the $\hbar^0 $ plus the
$\hbar^2 $ contributions. The difference between the semiclassical
expansion and the exact result is due to pure shell effects which are
comparable in magnitude to the $\hbar^2 $ contributions and cannot be
incorporated within a semiclassical treatment.

\section{Acknowledgements}
We thank X. Vi\~nas for drawing our attention to the Strutinsky average
method. One of us (E.R.A.) acknowledges the members of the NIKHEF-K Theory
Group for the hospitality extended to him while this work
has been completed.
J. Caro acknowledges the Spanish M.E.C. for a grant.
This work has been partially supported by the DGICYT under contract
PB92-0927 and the Junta de Andaluc\'\i a (Spain) as well as
the FOM and NWO (The Netherlands).

\newpage

\section*{Appendix. Semiclassical Expansion of the Functional
Determinant}
\label{sec-sem-ex}

In this appendix we briefly review a very powerful method to compute
a derivative expansion of the functional determinant,
which has been first suggested within Quantum Field Theory \cite{Ch86}
to compute
effective actions in the low energy limit. It only requires the operator
inside the determinant
to be a differential operator of elliptic type, i.e. of the form
\be
P^2 + U(X)
\ee
where $P_\mu=i\hbar\,\partial_\mu+A_\mu(X)$
is the generalized momentum operator containing in the
most general case a non-abelian gauge field $A_\mu$. For completeness we
just quote the main steps of the method presented in \cite{Ch86, Ca93}.
$U(X)$ is an
operator depending on the position operator $X_\mu$ and can have any internal
symmetry structure. The following general
relations are satisfied
\bea
\left[ X_\mu, X_\nu \right] & = & 0 \nonumber \\
\left[ P_\mu, X_\nu \right] & = & i \hbar \, \delta_{\mu \nu}
\nonumber \\
\left[ P_\mu, P_\nu \right] & = & i\, F_{\mu \nu } \nonumber \\
\left[ P_\mu, U \right] & = & i\, {\cal D}_\nu U
\eea
The Euclidean indices $\mu$ and $\nu$ run
from 1 to D (the Euclidean space dimension). The basic object to be
expanded is the determinant of the elliptic operator
\bea
W&=&\log\hbox{\rm Det}\,(P^2+U(X))\nonumber\\
& =&
\hbox{\rm Tr}\,\log(P^2+U(X)).
\eea
In general this is an ultraviolet divergent quantity which in principle
requires regularization. However, in the applications we will be
considering only the phase of the determinant which turns out to be
convergent, and hence no explicit ultraviolet regularization will be
introduced. In addition, the determinant is invariant under similarity
transformations. A particularly interesting one is given by the
translation operator in momentum space, which generates the identity
\begin{equation}
 W= \hbox{\rm Tr}\,\left(e^{\frac{i}{\hbar}kX}\log(P^2+U(X))
e^{-\frac{i}{\hbar}kX}
\right)=
\hbox{\rm Tr}\,\log\left((P_\mu+
k_\mu)^2+U(X)\right),
\end{equation}
where $k$ is an arbitrary c-numerical vector. Since this is valid for
any momentum $k$, one can average over all possible values, as follows
\begin{eqnarray}
W &=& \Omega^{-1}\int{d^Dk\over (2\pi\hbar)^D} \, \hbox{\rm Tr}\,\log
\left(\Delta^{-1}
+2kP+P^2\right)  \nonumber \\
&\Omega &= \int{d^Dk\over (2\pi\hbar)^D}\,,\hspace{1cm} \Delta(X)=
\left(k^2+U(X)\right)^{-1}.
\end{eqnarray}
It should be mentioned that
the averaging procedure corresponds to project onto the gauge
invariant subspace, and hence implicitly assumes a gauge invariant
regulator for $W$. This expression suggests an expansion in the
generalized momentum operator $P_\mu$ by use of the  formula
\be
W = \Omega^{-1}\int{d^Dk\over (2\pi\hbar)^D}\hbox{\rm Tr}\,\left(
\log(\Delta^{-1}) - \sum_{n=1}^\infty
{(-1)^n\over n}(\Delta(2kP+P^2))^n\right).\nonumber\\
\ee

Although the starting average defines a gauge invariant object, the
former expansion does not preserve gauge invariance explicitly,
i.e. term by term. The main
goal is to restore this invariance in an explicit manner. We will not
dwell upon the details which have been described at length
elsewhere \cite{Ca93}. The final result can be cast in the form
\begin{equation}
W=\int \frac{d^Dk\, d^Dx}{(2\pi\hbar)^D} \, \sum_{n=0}^\infty
\hbar^{2n}\,W_{2n}(x,k)\,.
\label{eq-app}
\end{equation}
For later reference we reproduce in table \ref{t-Chan} the general
results up to sixth order.
Note that the infinite volume factor $\Omega$ cancels in the final
result, as it should be.

\newpage

\section*{TABLE CAPTIONS}

\begin{enumerate}
\item  \label{t-den-bar}
Wigner-Kirkwood expansion of the valence contribution to the
baryonic density, $\langle \psi^\dagger(x) \psi(x)
\rangle_{\rm{val}}$,
in the relativistic case up to fourth order in $\hbar$ for $D=3$
dimensions. The definitions $\epsilon_F=\mu -V$,
$k_F = (\epsilon_F^2 - \Phi^2)^{1/2}$,
$x_F=\epsilon_F/k_F$ and $l_F=\log((\epsilon_F + k_F)/\Phi)$ have been used.
For more details sea main text and eq.~(\ref{sub}).

\item \label{t-den-is}
The same as table~\ref{t-den-bar} but for the scalar density,
$\langle {\overline \psi}(x)
\psi(x)\rangle_{\rm{val}}$.

\item \label{t-den-en}
The same as table~\ref{t-den-bar} but for the valence
contribution to the total fermionic energy density, $\langle
\psi^\dagger(x) ( - i \alpha \cdot \nabla + \beta \Phi(x) + V(x) )
\psi(x) \rangle_{\rm{val}}$.

\item \label{t-arm}
Wigner-Kirkwood expansion for the total energy (in MeV) of a
system of
$A$ nucleons in relativistic
harmonic potentials $V=\Phi - m=\frac{1}{4} m \omega^2 r^2$ where
$\hbar
\omega=41 A^{-1/3}$. WK$_n$ includes the semiclassical corrections up to
$\hbar^n$.
NP represents the non-perturbative determination of the chemical
potential, whereas P stands for the perturbative determination (see main
text). SA is the Strutinsky averaged total energy with Gaussian weight
factors. E$_{{\rm ex}}$ is the exact quantum-mechanical total energy. In
all cases the energy $m A$ has been subtracted. The numbers for $A$,
($A=Z+N$ and $Z=N$) represent the corresponding magic numbers for these
potentials.

\item \label{t-den-sea}
WK$\hbar^4$-corrections to the Dirac sea baryonic, scalar and total energy
densities respectively.

\item \label{t-fit}
Numerical fit of the parameters of the Woods-Saxon Potentials
(see eq.~(\ref{WS-pot})) to the self-consistent mean field $\phi$,
$V_0$, solutions in the Hartree valence approximation with 49 points.
$\sigma$ (in MeV) is the standard deviation between the Hartree
solutions and the Woods-Saxon potentials within the physically relevant
region $0 \le r \le 12$ fm. $U_o$ is given in MeV, and $R_o$ and $a_o$
are presented in fm.

\item \label{t-prop}
Comparison of the binding energy per nucleon, B/A (MeV), and
mean squared charge radius, m.s.c.r. (fm), for several closed shell
nuclei. $\sigma$-$\omega$ means the $\sigma$-$\omega$ self consistent
Hartree solutions with the parameter set specified in the text. WS stands
for the best fit of Woods-Saxon potentials to the corresponding self
consistent scalar and vector fields. For general information we also
quote the experimental values whenever they are accessible.

\item \label{t-WS-res}
Semiclassical expansion for the total fermionic energy in MeV
for several closed shell nuclei. The scalar an vector potentials are of
Woods-Saxon type fitted to the self consistent Hartree valence
potentials (see table~\ref{t-fit}). The chemical potential has been
always adjusted within the perturbative method (see
section~\ref{sec-reo}). WK$_n$ contains the corrections up to order
$\hbar^n$. SA stands for the Strutinsky averaged energy and WS
represents the quantum mechanical energy. In all cases the rest mass of
the constituent nucleons $M$ has been subtracted.

\item \label{t-Chan}
Values of the coefficients $W_{2 n}(x,k)$ as defined in
eq.~(\ref{eq-app}). The following convention has been used:
\begin{enumerate}
\item Given an object $Y_I$ with $I=(\alpha_1 \ldots \alpha_n)$ a set of
Euclidean indices we define $Y_{I \mu}$ to mean $- i [P_\mu, Y_I]$; for
instance
$
\Delta_{\mu \nu} = (- i)^2 [P_\nu , [P_\mu, \Delta]]
$ ; $F_{\mu \nu \rho} = - i [P_\mu, F_{\mu \nu}].$
\item Each term in the table means the half-sum of the term itself and
its mirror symmetric. The mirror symmetry is defined as
$P_\mu \to P_\mu^T = P_\mu$; $\Delta \to \Delta^T = \Delta$ and
$YZ \to (YZ)^T = Z Y.$ For instance $\Delta_\mu^T = - \Delta_\mu$;
$F_{\mu \nu}^T = - F_{\mu \nu}$; $F_{\mu \nu \rho}^T = F_{\mu \nu
\rho}.$
\item Notice the cyclic permutation symmetry.
\end{enumerate}

\end{enumerate}

\section*{FIGURE CAPTION}
\begin{enumerate}
\item \label{f-complex}
Contour in the radial complex plane used to evaluate radial
integrals in the classically allowed region (defined by the path $C_+$),
namely $\int_{C_+} = - \frac{1}{2} \int_{C_1}$. The branch cut starts at
the radial turning point $r_c(\lambda)$ defined in the text.
\end{enumerate}
\newpage

%\begin{table}[p]
\begin{center}
\newcommand{\nl}{\\}
$$
\begin{array}{|l|}\hline
\rho^{+}_0 = \frac{k_{F}^3}{3\, \pi^2\, \hbar^3}\, \Theta(\mu - V - \Phi)
\rule[-3mm]{0cm}{9mm}
\\\hline
\rho^{+}_2 = \frac{1}{24\, \pi^2\,\hbar} \Biggl\{
\Biggl.
\hfill
{{1\over {k_{F}}} \left( 2 - {{x_{F}}^2} \right)
        {{(\nabla \Phi )}^2}}
 \hfill + \hfill   {{2\over {\Phi }} x_{F} \left( 3 - {{x_{F}}^2} \right)
        (\nabla_{i} \Phi ) (\nabla_{i} V)}
\rule{0cm}{8mm}\nl
 + {{1\over {k_{F}}} \left( 3 - {{x_{F}}^2} \right)
        {{(\nabla V)}^2}}
\hfill - \hfill {{2\over{k_{F}}} \Phi  (\nabla^2 \Phi )}
\hfill - \hfill
       2 \left( x_{F} + 2 l_{F} \right)
(\nabla^2 V)
\hfill
\Biggl. \Biggr\}\,\Theta(\mu -V -\Phi)
\rule[-5mm]{0cm}{1mm}
\\\hline
\rho^{+}_{4} = \frac{\hbar}{24 \, \pi^2} \Biggl\{ \Biggr.
\hfill
    {{1\over
      {10  {{\, k_{F}}^3}}}  \, \Phi   (\nabla^4 \Phi  )}
 \hfill + \hfill  {{1\over
      {10  {{\, \Phi  }^2}}}  \, x_{F}
         \left( -5 + {{\, x_{F}}^2} \right)
        \, (\nabla^4 V )}
\rule{0cm}{8mm}\nl
 + {{1\over {20  {{\, k_{F}}^3}}}
         \left( -5 + 4  {{\, x_{F}}^2} \right)
        {{\, (\nabla_{i}\nabla_{j}\Phi  )}^2}}
 \hfill + \hfill
    {{1\over {60  {{\, k_{F}}^3}}}   \left( -19 + 12  {{\, x_{F}}^2} \right)
    {{\, (\nabla_{i}\nabla_{j} V )}^2}} \nl
 +
      {{1\over {30  {{\, \Phi  }^3}}}  \, x_{F} \left( 45 - 35  {{\, x_{F}}^2}
 \hfill + \hfill
          12  {{\, x_{F}}^4} \right)   \, (\nabla_{i}\nabla_{j}\Phi  )
         (\nabla_{i}\nabla_{j} V )} \nl
+ {{3\over
        {10  {{\, \Phi  }^3}}}  \, x_{F}
         \left( 5 - 5  {{\, x_{F}}^2} + 2  {{\, x_{F}}^4} \right)
        \, (\nabla_{i} V )  (\nabla_{i}\nabla^2\Phi  )}\nl
 +
    {{1\over
      {10  {{\, k_{F}}^3}}}   \left( -7 + 6  {{\, x_{F}}^2} \right)
     \, (\nabla_{i}\Phi  )
         (\nabla_{i}\nabla^2\Phi  )}
 \hfill + \hfill
   {{1\over {5  {{\, k_{F}}^3}}
      }
         \left( -5 + 3  {{\, x_{F}}^2} \right)   \, (\nabla_{i} V )
         (\nabla_{i}\nabla^2 V )} \nl
 + {{1\over
        {5  {{\, \Phi  }^3}}}  \, x_{F}
         \left( 15 - 10  {{\, x_{F}}^2} + 3  {{\, x_{F}}^4} \right)
        \, (\nabla_{i}\nabla^2 V )  (\nabla_{i}\Phi  )}  \nl
 + {{1\over
      {10  {{\, k_{F}}^5}}}  \, \Phi
         \left( -14 + 15  {{\, x_{F}}^2} \right)   \, (\nabla_{i}\Phi  )
         (\nabla_{i}\nabla_{j}\Phi  )  (\nabla_{j}\Phi  )} \nl
 + {{1\over
      {10  {{\, \Phi  }^4}}}  \, x_{F}
         \left( -75 + 125  {{\, x_{F}}^2} - 102  {{\, x_{F}}^4} +
          30  {{\, x_{F}}^6} \right)   \, (\nabla_{i} V )
         (\nabla_{i}\nabla_{j}\Phi  )   (\nabla_{j}\Phi  )} \nl
 + {{1 \over
      {30  {{\, \Phi  }^4}}} \, x_{F}
        \left( -135 + 205  {{\, x_{F}}^2} - 159  {{\, x_{F}}^4} +
          45  {{\, x_{F}}^6} \right)   \, (\nabla_{i}\Phi  )
         (\nabla_{i}\nabla_{j} V )   (\nabla_{j}\Phi  )} \nl
 + {{3 \over
      {  {{\, k_{F}}^7}}}  {{\, \Phi  }^3}  \, (\nabla_{i} V )
         (\nabla_{i}\nabla_{j} V )  (\nabla_{j}\Phi  )}
 \hfill + \hfill
     {{1 \over
      {80  {{\, k_{F}}^5}}}
         \left( 96 - 280  {{\, x_{F}}^2} + 175  {{\, x_{F}}^4} \right)
        {{\, (\nabla \Phi  )}^4}}  \nl
+ {{1 \over
      {60  {{\, \Phi  }^5}}} \, x_{F}
         \left( 900 - 2455  {{\, x_{F}}^2} + 3351  {{\, x_{F}}^4} -
          2145  {{\, x_{F}}^6} + 525  {{\, x_{F}}^8} \right)
     \, (\nabla_{i}\Phi  )(\nabla_{i} V )  {{(\nabla \Phi  )}^2}}\nl
 + {{1\over
      {8  {{\, k_{F}}^5}}}
         \left( 18 - 55  {{\, x_{F}}^2} + 35  {{\, x_{F}}^4} \right)
        {{\, (\nabla V )}^2}  {{ (\nabla \Phi  )}^2}} \nl
 + {{1 \over {12  {{\, \Phi  }^4}}} \, x_{F}
         \left( -54 + 73  {{\, x_{F}}^2} - 54  {{\, x_{F}}^4} +
          15  {{\, x_{F}}^6} \right)   \, (\nabla^2 V )
        {{ (\nabla \Phi  )}^2}} \nl
 +
    {{1\over
     {10  {{\, k_{F}}^5}}}  \, \Phi     \left( -13 + 15  {{\, x_{F}}^2} \right)
        \, (\nabla_{i} V )   (\nabla_{i}\nabla_{j}\Phi  ) (\nabla_{j} V )} \nl
 +
    {{1\over
      {30  {{\, \Phi  }^4}}}  \, x_{F}   \left( -90 + 185  {{\, x_{F}}^2} -
          156  {{\, x_{F}}^4} + 45  {{\, x_{F}}^6} \right)   \, (\nabla_{i} V )
        (\nabla_{i}\nabla_{j} V )   (\nabla_{j} V )} \nl
 + {{1 \over {20  {{\, k_{F}}^5}}}
         \left( 89 - 275  {{\, x_{F}}^2} + 175  {{\, x_{F}}^4} \right)
        \, (\nabla_{i}\Phi  )  (\nabla_{i} V )  (\nabla_{j}\Phi  )
         (\nabla_{j} V )}\nl
 +
    {{1\over {60  {{\, \Phi  }^5}}}  \, x_{F} \left( 675 - 2270{{\, x_{F}}^2}+
      3264  {{\, x_{F}}^4} - 2130  {{\, x_{F}}^6} + 525  {{\, x_{F}}^8} \right)
     {{\, (\nabla V )}^2} (\nabla_{i}\Phi  ) (\nabla_{i} V )}
        \nl
 +
    {{1\over
    {10  {{\, k_{F}}^5}}}  \, \Phi  \left( -26 + 25  {{\, x_{F}}^2} \right)
        \, (\nabla^2 V )   (\nabla_{i}\Phi  ) (\nabla_{i} V )} \nl
 +
    {{1 \over
    {80  {{\, k_{F}}^5}}}  \left( 63 - 270  {{\, x_{F}}^2} +
     175  {{\, x_{F}}^4} \right)
    {{\, (\nabla V )}^4}} \nl
 + {{1  \over {20  {{\, k_{F}}^5}}}\, \Phi
         \left( -18 + 25  {{\, x_{F}}^2} \right)  {{\, (\nabla \Phi  )}^2}
        (\nabla^2 \Phi  )} \nl
 +
    {{1\over {30  {{\, \Phi  }^4}}}   \, x_{F}  \left( -135 +
         265  {{\, x_{F}}^2} -
          237  {{\, x_{F}}^4} + 75  {{\, x_{F}}^6} \right)
        \, (\nabla_{i}\Phi  )  (\nabla_{i} V )
        (\nabla^2 \Phi  )} \nl
 +
    {{1 \over
      {4  {{\, k_{F}}^5}}}  \, \Phi  \left( -3 + 5  {{\, x_{F}}^2} \right)
        {{\, (\nabla V )}^2} (\nabla^2 \Phi  )}
  \hfill + \hfill {{1  \over
      {20  {{\, k_{F}}^3}}}
        \left( -3 + 5  {{\, x_{F}}^2} \right)
        \, (\nabla^2 \Phi  )^2}\nl
 +
    {{1\over {6  {{\, \Phi  }^3}}}  \, x_{F} \left( 9 - 8  {{\, x_{F}}^2} +
          3  {{\, x_{F}}^4} \right)   \, (\nabla^2 V )
         (\nabla^2 \Phi  )} \nl
 +
    {{1 \over
      {60  {{\, \Phi  }^4}}}  \, x_{F}  \left( -135 + 305  {{\, x_{F}}^2} -
          261  {{\, x_{F}}^4} + 75  {{\, x_{F}}^6} \right)
        {{\, (\nabla V )}^2} (\nabla^2 V )}\nl
+ {{1  \over
      {60  {{\, k_{F}}^3}}}
        \left( -26 + 15  {{\, x_{F}}^2} \right)
        \, (\nabla^2 V )^2 }
 \hfill + \hfill  {1\over { \hbar^4 \, k_{F}}} \, {{F_{ij}^2}}
 \hfill
 \Biggl. \Biggr\} \, \Theta(\mu - V - \Phi)
\rule[-5mm]{0cm}{1mm}\\  \hline
\end{array}
$$
\\
Table \ref{t-den-bar}
\end{center}
%\end{table}

\newpage

%\begin{table}[p]
\begin{center}
\newcommand{\nl}{\\}
$$
\begin{array}{|l|}\hline
{\rho^+_s}_{0} = \frac{1}{2 \, \pi^2\, \hbar^3} \left\{
\epsilon_F \, k_F \, \Phi
- \Phi^3 \,l_F
\right\}
\, \Theta(\mu-V-\Phi)
\rule[-3mm]{0cm}{9mm}
\\\hline
{\rho^+_s}_{2} = \frac{1}{24 \, \pi^2\, \hbar} \Biggl\{ \Biggr.
\hfill {-{ 1\over {\Phi }}\,x_{F}\left( 2 + {{x_{F}}^2} \right)\,
          {{(\nabla \Phi )}^2}  }
\hfill - \hfill
    {{2\over {k_{F}}}\,\Phi (\nabla^2  V )}
\hfill - \hfill
    {{2\over {k_{F}}}\left( 2 + {{x_{F}}^2} \right) \,(\nabla_{i} \Phi )
        (\nabla_{i}  V )}
\rule{0cm}{8mm}
\nl
 +
    {{1\over
      {\Phi }}\,x_{F}\,\left( 1 + {{x_{F}}^2} \right) \,
        {{(\nabla  V )}^2}}
\hfill + \hfill
{    2 \left( - x_{F}  + 3 \,l_{F} \right) \ (\nabla^2 \Phi ) }
\hfill \Biggl. \Biggr\} \, \Theta(\mu-V-\Phi)
\rule[-5mm]{0cm}{1mm}
\\\hline
%
% rho4
%
{\rho^+_s}_{4} = \frac{\hbar}{24\, \pi^2} \Biggl\{ \Biggr.
    {{1\over
      {10\,{{\Phi }^2}}}\, x_{F} \left( 2 + {{ x_{F} }^2} \right) \,
        (\nabla^4 \Phi )}
\hfill + \hfill {{1\over
      {10\,{{ k_{F} }^3}}}\,\Phi \,
        (\nabla^4  V )}
\rule{0cm}{8mm}
\nl
+ {{1\over {60\,{{\Phi }^3}}}\, x_{F}
        \left( -21 - 13\,{{ x_{F} }^2} + 12\,{{ x_{F} }^4} \right) \,
        {{(\nabla_{i} \nabla_{j} \Phi )}^2}}
\hfill + \hfill
    {{5\over
      {4\,{{ k_{F} }^5}}}\,\Phi \,{{ x_{F} }^2}\,
        (\nabla^2  V ){{(\nabla \Phi )}^2}}
\nl
+
    {{1\over
        {10\,{{ k_{F} }^3}}}\left( 1 + 4\,{{ x_{F} }^2} \right) \,
        (\nabla_{i} \nabla_{j} \Phi )(\nabla_{i} \nabla_{j}  V )}
\hfill + \hfill
    {{1\over
      {20\,{{ k_{F} }^5}}}\,\Phi \left( 3 + 25\,{{ x_{F} }^2} \right) \,
        {{(\nabla  V )}^2}(\nabla^2  V )}
\nl
+
    {{1\over
 {60\,{{\Phi }^3}}}\, x_{F} \left( -9 - 17\,{{ x_{F} }^2} +
          12\,{{ x_{F} }^4} \right) \,{{(\nabla_{i} \nabla_{j}  V )}^2}}
\hfill + \hfill
    {{1\over {2\,{{ k_{F} }^3}}}\,{{ x_{F} }^2}\,(\nabla^2  V )
        (\nabla^2 \Phi )}
\nl
+
    {{3\over {5\,{{\Phi }^3}}}\, x_{F} \,\left( -2 - {{ x_{F} }^2}
      + {{ x_{F} }^4} \right)
    \, (\nabla_{i} \Phi )(\nabla_{i} \nabla^2 \Phi )
        }
\nl
+
    {{1\over
      {5\,{{ k_{F} }^3}}} \left( 2 + 3\,{{ x_{F} }^2} \right)
	\,(\nabla_{i}  V ) (\nabla_{i} \nabla^2 \Phi )}
\hfill + \hfill
    {{1\over
        {10\,{{ k_{F} }^3}}}\,
        \left( -1 + 6\,{{ x_{F} }^2} \right) \,
        (\nabla_{i} \nabla^2  V )(\nabla_{i} \Phi )}
\nl
+ {{3\over
      {10\,{{\Phi }^3}}}\, x_{F}
        \left( -1 - 3\,{{ x_{F} }^2} + 2\,{{ x_{F} }^4} \right) \,
        (\nabla_{i}  V )(\nabla_{i} \nabla^2  V )}
\nl
+
    {{1\over {30\,{{\Phi }^4}}}\, x_{F} \left( 60 + 35\,{{ x_{F} }^2} -
          96\,{{ x_{F} }^4} + 45\,{{ x_{F} }^6} \right) \,
        (\nabla_{i} \Phi )(\nabla_{i} \nabla_{j} \Phi )
        (\nabla_{j} \Phi )}
\nl
+
    {{3\over {5\,{{ k_{F} }^5}}}\,\Phi \left( 2 + 5\,{{ x_{F} }^2} \right) \,
        (\nabla_{i}  V )(\nabla_{i} \nabla_{j} \Phi )\,
        (\nabla_{j} \Phi )}
\nl
+
    {{1\over {10\,{{ k_{F} }^5}}}\,\Phi\left( 4 + 15\,{{ x_{F} }^2} \right) \,
        (\nabla_{i} \Phi )(\nabla_{i} \nabla_{j}  V )
        (\nabla_{j} \Phi )}
\nl
+
    {{1\over
      {10\,{{\Phi }^4}}}\, x_{F} \left( 15 + 35\,{{ x_{F} }^2} -
          66\,{{ x_{F} }^4} + 30\,{{ x_{F} }^6} \right)
	(\nabla_{i}  V )
        (\nabla_{i} \nabla_{j}  V )(\nabla_{j} \Phi )}
\nl
+
    {{1\over
      {80\,{{\Phi }^5}}}\, x_{F}
        \left( -160 - 80\,{{ x_{F} }^2} + 507\,{{ x_{F} }^4} -
          530\,{{ x_{F} }^6} + 175\,{{ x_{F} }^8} \right) \,
        {{(\nabla \Phi )}^4}}
\nl
+
    {{1\over
      {4\,{{ k_{F} }^5}}}
        \left( -8 - 18\,{{ x_{F} }^2} + 35\,{{ x_{F} }^4} \right) \,
        (\nabla_{i} \Phi )(\nabla_{i}  V ){{(\nabla \Phi )}^2}}
\nl
+
    {{1\over
      {24\,{{\Phi }^5}}}\, x_{F}
        \left( -30 - 85\,{{ x_{F} }^2} + 309\,{{ x_{F} }^4} -
          315\,{{ x_{F} }^6} + 105\,{{ x_{F} }^8} \right) \,
        {{(\nabla  V )}^2}{{(\nabla \Phi )}^2}}
\nl
+
    {{1\over
      {30\,{{\Phi }^4}}}\, x_{F}
        \left( 30 + 40\,{{ x_{F} }^2} - 93\,{{ x_{F} }^4} +
          45\,{{ x_{F} }^6} \right) \,(\nabla_{i}  V )
        (\nabla_{i} \nabla_{j} \Phi )(\nabla_{j}  V )}
\nl
+
    {{1\over
      {2\,{{ k_{F} }^5}}}\,\Phi
        \left( 1 + 3\,{{ x_{F} }^2} \right) \,(\nabla_{i}  V )
        (\nabla_{i} \nabla_{j}  V )(\nabla_{j}  V )}
\nl
+
    {{1\over {60\,{{\Phi }^5}}}\, x_{F}
        \left( -165 - 415\,{{ x_{F} }^2} + 1542\,{{ x_{F} }^4} -
          1575\,{{ x_{F} }^6} + 525\,{{ x_{F} }^8} \right) \,
        \left((\nabla_{i} \Phi )(\nabla_{i}  V )\right)^2}
\nl
+
    {{1\over
      {4\,{{ k_{F} }^5}}}\left( -6 - 17\,{{ x_{F} }^2} +
             35\,{{ x_{F} }^4} \right) \,
        {{(\nabla  V )}^2}(\nabla_{j} \Phi )(\nabla_{j}  V )}
\nl
+
    {{1\over
	{30\,{{\Phi }^4}}}\,{{ x_{F} }^3}
        \left( 130 - 183\,{{ x_{F} }^2} + 75\,{{ x_{F} }^4} \right) \,
        (\nabla^2 V )(\nabla_{j} \Phi)(\nabla_{j}  V )}
\nl
+
    {{1\over
      {240\,{{\Phi }^5}}}\, x_{F} \left( -135 - 400\,{{ x_{F} }^2} +
          1506\,{{ x_{F} }^4} - 1560\,{{ x_{F} }^6} + 525\,{{ x_{F} }^8}
        \right) \,{{(\nabla  V )}^4}}
\nl
+
    {{1\over {60\,{{\Phi }^4}}}\, x_{F}
        \left( 120 + 50\,{{ x_{F} }^2} - 159\,{{ x_{F} }^4} +
          75\,{{ x_{F} }^6} \right) \,{{(\nabla \Phi )}^2}\,
        (\nabla^2 \Phi )}
\nl
+
  {{1\over {10\,{{ k_{F} }^5}}}\,\Phi \,\left( 11 + 25\,{{x_{F}}^2}\right)\,
        (\nabla_{i} \Phi )\,(\nabla_{i}  V )\,
        (\nabla^2  \Phi )}
\nl
+
    {{1\over
      {12\,{{\Phi }^4}}}\, x_{F} \,\left( 6 + 13\,{{ x_{F} }^2} -
          30\,{{ x_{F} }^4} + 15\,{{ x_{F} }^6} \right) \,
        {{(\nabla  V )}^2}(\nabla^2 \Phi )}
\nl
+
    {{1\over {60\,{{\Phi }^3}}}\, x_{F} \,
        \left( -33 - 14\,{{ x_{F} }^2} + 15\,{{ x_{F} }^4} \right) \,
        (\nabla^2 \Phi )(\nabla^2 \Phi )}
\nl
+ {{1\over
      {60\,{{\Phi}^3}}}\, x_{F}
        \left( 18 - 31\,{{ x_{F} }^2} + 15\,{{ x_{F} }^4} \right) \,
        (\nabla^2  V )^2}
\hfill + \hfill
    {{1\over {\hbar^4\,\Phi }}\, x_{F} \,{{F_{ij}^2}}}
\Biggl. \Bigg\} \, \Theta(\mu-V-\Phi)
\rule[-5mm]{0cm}{1mm}
\\\hline
\end{array}
$$
\\
Table \ref{t-den-is}
\end{center}
%\end{table}

\newpage

%\begin{table}[p]
\begin{center}
\newcommand{\nl}{\\}
$$
\begin{array}{|l|}\hline
{\rho^+_E}_0 = \frac{1}{8 \, \pi^2\,\hbar^3} \left\{
\epsilon_{F}^3 \, k_{F}
+
 \epsilon_{F} \, k_{F}^3
-
\Phi^4 \, l_{F}
\right\} \Theta(\mu - V - \Phi) + \rho^+_0 \, V
\rule[-3mm]{0cm}{9mm}
\\\hline
{\rho^+_E}_2 =
\frac{1}{24\,\pi^2\,\hbar} \Biggl\{ \Biggl.
\hfill
{2\,\Phi}
        \left( - x_{F} +  l_{F} \right)
\, (\nabla^2 \Phi)
\hfill - \hfill
{2\,{\Phi^3 \over {{k_{F}}^3}}
\, (\nabla_{i} \Phi) (\nabla_{i} V)}
\hfill - \hfill
{2\,{k_{F}} \left( 1 + {{x_{F}}^2} \right)
        \,(\nabla^2 V)}
\rule{0cm}{8mm}
\nl
+  \left( x_{F} - {{x_{F}}^3} - l_{F} \right)
{{{(\nabla  \Phi) }^2} }
\hfill + \hfill
   \left( 2 x_{F} - {{x_{F}}^3} - 2 l_{F} \right)
{{(\nabla V)}^2}
\hfill
\Biggl. \Biggr\} \Theta(\mu - V - \Phi)
\hfill+ \hfill
\rho^+_2 \, V
\rule[-5mm]{0cm}{1mm}
\\\hline
{\rho^+_E}_{4} =
\frac{\hbar}{24 \, \pi^2} \, \Biggl\{ \Biggr.
\hfill
{{1 \over
      {10  \, \Phi }} {{  \, x_{F} }^3}
         \, (\nabla^4 \Phi )}
\hfill + \hfill
{{1 \over
      {10    \, k_{F}}} \left( -3 + {{ x_{F} }^2} \right)
         \, (\nabla^4 V)}
\rule{0cm}{8mm}
\nl
+ {{1 \over {60  {{\, \Phi }^2}}} {{  \, x_{F} }^3}
         \left( -23 + 12  {{  \, x_{F} }^2} \right)
        {{ \, (\nabla_{i} \nabla_{j} \Phi )}^2}}
\hfill + \hfill
    {{1 \over
        {30  {{  \, k_{F}}^3}}}  \, \Phi \left( -7 +
              12  {{  \, x_{F} }^2} \right)
         \, (\nabla_{i} \nabla_{j} \Phi )   \, (\nabla_{i} \nabla_{j} V)}
\nl
+
    {{1\over {20  {{\, \Phi }^2}}}  {{  \, x_{F} }^3} \left( -9 +
                   4  {{  \, x_{F} }^2} \right)
        {{ \, (\nabla_{i} \nabla_{j} V)}^2}}
\hfill + \hfill
    {{1 \over
 {10  {{\, \Phi }^2}}} {{  \, x_{F} }^3} \left( -11 +
             6  {{  \, x_{F} }^2} \right)
         \, (\nabla_{i} \Phi )  (\nabla_{i} \nabla^2 \Phi )}
\nl
+
    {{1 \over
        {10  {{  \, k_{F}}^3}}} \, \Phi \left( -1 +
       6  {{  \, x_{F} }^2} \right)
         \, (\nabla_{i} V)  (\nabla_{i} \nabla^2 \Phi )}
\hfill + \hfill
{{3 \over
        {5  {{  \, k_{F}}^5}}} {{\, \Phi }^3}
         \, (\nabla_{i} \nabla^2 V) (\nabla_{i} \Phi )}
\nl
+
    {{1 \over
      {5 {{\, \Phi }^2}}} {{  \, x_{F} }^3}  \, \left( -7 +
         3  {{  \, x_{F} }^2} \right)
         \, (\nabla_{i} V) (\nabla_{i} \nabla^2 V)}
\nl
+
    {{1\over {30  {{\, \Phi }^3}}}  {{  \, x_{F} }^3}
         \left( 100 - 123  {{  \, x_{F} }^2} + 45  {{  \, x_{F} }^4} \right)
         \, (\nabla_{i} \Phi )  (\nabla_{i} \nabla_{j} \Phi )
          (\nabla_{j} \Phi )}
\nl
+
    {{1 \over {10 {{\, k_{F}}^3}}}\left( 1 - 36  {{  \, x_{F} }^2} +
          30  {{  \, x_{F} }^4} \right)
         \, (\nabla_{i} V) (\nabla_{i} \nabla_{j} \Phi )
         (\nabla_{j} \Phi )}
\nl
+
    {{1 \over {6  {{  \, k_{F}}^3}}} \left( 1 - 12  {{  \, x_{F} }^2} + 9
          {{  \, x_{F} }^4} \right)
         \, (\nabla_{i} \Phi ) (\nabla_{i} \nabla_{j} V)
         \, (\nabla_{j} \Phi )}
\nl
+
    {{1 \over
 {5 {{\Phi }^3}}} {{  \, x_{F} }^3}
        \left( 35 - 42  {{  \, x_{F} }^2} + 15  {{  \, x_{F} }^4} \right)
         \, (\nabla_{i} V) (\nabla_{i} \nabla_{j} V) (\nabla_{j} \Phi )}  \nl
+
    {{1 \over {48  {{\, \Phi }^4}}} {{  \, x_{F} }^3}
        \left( -208 + 444  {{  \, x_{F} }^2} - 363  {{  \, x_{F} }^4} +
          105  {{  \, x_{F} }^6} \right)   {{ \, (\nabla \Phi )}^2}
        {{ (\nabla \Phi )}^2}}
\nl
+
    {{1\over {20  {{  \, k_{F}}^5}}}  \, \Phi \left( -6 -
          165  {{  \, x_{F} }^2} +
          175  {{  \, x_{F} }^4} \right)
      \, (\nabla_{i} \Phi )  (\nabla_{i} V)
        {{(\nabla \Phi )}^2}}  \nl
+
    {{5  \over {24  {{\, \Phi }^4}}}  {{  \, x_{F} }^3}
        \, \left( -40 + 87  {{  \, x_{F} }^2} - 72  {{  \, x_{F} }^4} +
          21  {{  \, x_{F} }^6} \right)   {{ \, (\nabla V)}^2}
        {{(\nabla \Phi )}^2}}
\nl
+
    {{1 \over
      {12 {{  \, k_{F}}^3}}} \left( 4 - 21  {{  \, x_{F} }^2} +
          15  {{  \, x_{F} }^4} \right)
         \, (\nabla^2 V)  {{(\nabla \Phi )}^2}}
\nl
 + {{1 \over
     {6 {{\, \Phi }^3}}} {{  \, x_{F} }^3}
        \left( 19 - 24  {{  \, x_{F} }^2} + 9  {{  \, x_{F} }^4} \right)
         \, (\nabla_{i} V) (\nabla_{i} \nabla_{j} \Phi ) (\nabla_{j} V)}
\nl
+
    {{1 \over
      {30  {{  \, k_{F}}^3}}} \left( -8 - 57  {{  \, x_{F} }^2} +
          45  {{  \, x_{F} }^4} \right)
         \, (\nabla_{i} V) (\nabla_{i} \nabla_{j} V) (\nabla_{j} V)}
\nl
 + {{1 \over {60  {{\, \Phi }^4}}} {{  \, x_{F} }^3}
        \left( -995 + 2172  {{  \, x_{F} }^2} - 1800  {{  \, x_{F} }^4} +
          525  {{  \, x_{F} }^6} \right)
         \, \left((\nabla_{i} \Phi ) (\nabla_{i} V)\right)^2 }
\nl
 +
    {{1\over {20  {{  \, k_{F}}^5}}}  \, \Phi \left( -19 -
            160  {{  \, x_{F} }^2} +
          175  {{  \, x_{F} }^4} \right)   {{ \, (\nabla V)}^2}
         (\nabla_{i} \Phi ) (\nabla_{i} V)}
\nl
 +
    {{1 \over
     {10  {{\, \Phi }^3}}} {{  \, x_{F} }^3}
        \left( 60 - 71  {{  \, x_{F} }^2} + 25  {{  \, x_{F} }^4} \right)
         \, (\nabla^2 V) (\nabla_{i} \Phi ) (\nabla_{i} V)}
\nl
 +
    {{1 \over {80  {{\, \Phi }^4}}} {{  \, x_{F} }^3}
        \left( -285 + 689  {{  \, x_{F} }^2} - 595  {{  \, x_{F} }^4} +
          175  {{  \, x_{F} }^6} \right) {{ \, (\nabla V)}^4}
        }
\nl
 +
    {{1 \over
      {60  {{\, \Phi }^3}}} {{  \, x_{F} }^3}
         \left( 140 - 189  {{  \, x_{F} }^2} + 75  {{  \, x_{F} }^4} \right)
        {{ \, (\nabla \Phi )}^2}  (\nabla^2 \Phi )}
\nl
 + {{1  \over {10  {{  \, k_{F}}^3}}}
         \left( -1 - 24  {{  \, x_{F} }^2} + 25  {{  \, x_{F} }^4} \right)
         \, (\nabla_{i} \Phi )  (\nabla_{i} V)
         (\nabla^2 \Phi )}
\nl
 +
    {{1 \over
      {12  {{\, \Phi }^3}}}  {{  \, x_{F} }^3}
        \left( 25 - 36  {{  \, x_{F} }^2} + 15  {{  \, x_{F} }^4} \right)
        {{ \, (\nabla V)}^2} (\nabla^2 \Phi )}
\nl
 + {{1  \over
 {60  {{\, \Phi }^2}}}{{  \, x_{F} }^3}
         \left( -19 + 15  {{  \, x_{F} }^2} \right)
         \, (\nabla^2 \Phi )^2}
\hfill + \hfill
    {{1\over
        {6  {{  \, k_{F}}^3}}}   \, \Phi  \left( -1 +
         3  {{  \, x_{F} }^2} \right)
         \, (\nabla^2 V) (\nabla^2 \Phi )}
\nl
 +
    {{1 \over
      {60  {{  \, k_{F}}^3}}} \left( -19 - 96  {{  \, x_{F} }^2} +
         75  {{  \, x_{F} }^4} \right)
        {{ \, (\nabla V)}^2} (\nabla^2 V)}
\nl
 + {{1 \over
      {20  {{\, \Phi }^2}}}  {{  \, x_{F} }^3}
        \left( -12 + 5  {{  \, x_{F} }^2} \right)
         \, (\nabla^2 V)^2}
\hfill + \hfill
{1\over {\hbar^4}}   \left(x_{F}  - l_{F} \right) \, {{F_{ij}}^2}
\hfill
\Biggl. \Biggr\} \Theta(\mu-V-\Phi)
\hfill + \hfill
\rho^+_{4} \, V
\rule[-5mm]{0cm}{1mm}
\\\hline
\end{array}
$$
\\
Table \ref{t-den-en}
\end{center}
%\end{table}

\newpage

%\begin{table}[ht]
\newcommand{\col}[2]{\begin{tabular}{r}#1\\#2\end{tabular}}
\newcommand{\p}[1]{\phantom{#1}}
\begin{center}
\begin{tabular}{|r|rrrr|*{2}{r@{.}l|}}\hline
 \multicolumn{1}{|c|}{A} & & \multicolumn{1}{c}{WK$_0$} &
 \multicolumn{1}{c}{WK$_2$} & \multicolumn{1}{c}{WK$_4$} &
\multicolumn{2}{|c|}{SA}& \multicolumn{2}{|c|}{E$_{\rm{ex}}$} \\\hline
4   & \col{\rm{NP:}}{\rm{P:}}   &   \col{139.428}{139.428}
                 &   \col{162.018}{160.405}
                 &   \col{161.104}{161.066}
    &   161&1
    &   153&412  \\\hline
16  & \col{\rm{NP:}}{\rm{P:}}    &   \col{557.711}{557.711}
                 &   \col{592.013}{591.011}
                 &   \col{591.437}{591.427}
    &   591&44
    &   579&979  \\\hline
40  & \col{\rm{NP:}}{\rm{P:}}    &    \col{1394.28\p{0}}{1394.28\p{0}}
                 &    \col{1440.21\p{0}}{1439.47\p{0}}
                 &    \col{1439.78\p{0}}{1439.78\p{0}}
    &   1439&78
    &   1424&52  \\\hline
80  &\col{\rm{NP:}}{\rm{P:}}    &   \col{2788.56\p{0}}{2788.56\p{0}}
                 &   \col{2846.08\p{0}}{2845.50\p{0}}
                 &   \col{2845.74\p{0}}{2845.74\p{0}}
    &   2845&74
    &   2826&66  \\\hline
140 & \col{\rm{NP:}}{\rm{P:}}    &    \col{4879.97\p{0}}{4879.97\p{0}}
                 &    \col{4949.07\p{0}}{4948.59\p{0}}
                 &    \col{4948.79\p{0}}{4848.79\p{0}}
    &   4948&80
    &   4925&89  \\\hline
224 & \col{\rm{NP:}}{\rm{P:}}    &   \col{7807.96\p{0}}{7807.96\p{0}}
                 &   \col{7888.63\p{0}}{7888.21\p{0}}
                 &   \col{7888.39\p{0}}{7888.39\p{0}}
    &   7888&39
    &   7861&67\\\hline
\end{tabular}

Table \ref{t-arm}
\end{center}
%\end{table}

\newpage

%\begin{table}[p]
\begin{center}
\newcommand{\nl}{\\}
$$
\begin{array}{|l|}\hline
\rho^{\rm sea}_{4} = \frac{\hbar}{24 \, \pi^2} \Biggl\{ \Biggr.
{{2\over
      {5 }}
        \, \Phi^{-2} (\nabla^4 V )}
\rule{0cm}{8mm}
\hfill
-
\hfill
      {{11\over 15}
     \, \Phi^{-3}(\nabla_{i}\nabla_{j}\Phi  )
         (\nabla_{i}\nabla_{j} V )}
\hfill
-
\hfill
{{3\over
        5  }
        \, \Phi^{-3}(\nabla_{i} V )  (\nabla_{i}\nabla^2\Phi  )}\nl
 - {{8\over 5 }
        \, \Phi^{-3} (\nabla_{i}\nabla^2 V )  (\nabla_{i}\Phi  )}
\hfill
 +
\hfill
{{11\over 5}
           \, \Phi^{-4}(\nabla_{i} V )
         (\nabla_{i}\nabla_{j}\Phi  )   (\nabla_{j}\Phi  )}
\hfill +
\hfill
  {{22 \over 15}
             \, \Phi^{-4} (\nabla_{i}\Phi  )
         (\nabla_{i}\nabla_{j} V )   (\nabla_{j}\Phi  )} \nl
- {{44 \over
      15}
     \, \Phi^{-5} (\nabla_{i}\Phi  )(\nabla_{i} V )  {{(\nabla \Phi  )}^2}}
\hfill
 +
\hfill
{{5 \over 3}
     \, \Phi^{-4} (\nabla^2 V )  {{(\nabla \Phi  )}^2}}
\hfill
 +
\hfill
    {{8\over
      15}
             \, \Phi^{-4} (\nabla_{i} V )
        (\nabla_{i}\nabla_{j} V )   (\nabla_{j} V )} \nl
 -
    {{16\over 15}
     {{\, \Phi^{-5} (\nabla V )}^2} (\nabla_{i}\Phi  ) (\nabla_{i} V )}

\hfill
 +
\hfill
    {{16\over 15}
        \, \Phi^{-4} (\nabla_{i}\Phi  )  (\nabla_{i} V )
        (\nabla^2 \Phi  )}
\hfill
 -
\hfill
    {{2\over 3}
          \, \Phi^{-3}(\nabla^2 V )
         (\nabla^2 \Phi  )} \nl
 +
    {{4 \over
      15}
        {{\, \Phi^{-4} (\nabla V )}^2} (\nabla^2 V )}
 \Biggl. \Biggr\}
\hfill
\rule[-5mm]{0cm}{1mm}\\  \hline
% rho4s vac
%
\rho^{{\rm sea}}_{s4} = \frac{\hbar}{24\, \pi^2} \Biggl\{ \Biggr.
-
    {{3\over
      10}
      \, \Phi^{-2} (\nabla^4  \Phi )}
\rule{0cm}{8mm}

\hfill
+
\hfill
 {{11\over 30}
         \,\Phi^{-3}{{(\nabla_{i} \nabla_{j} \Phi )}^2}}
\hfill
+
\hfill
    {{7\over
   30}
           \,\Phi^{-3}{{(\nabla_{i} \nabla_{j}  V )}^2}} \nl
+
    {{6\over 5}
      \, \Phi^{-3}(\nabla_{i} \Phi )(\nabla_{i} \nabla^2 \Phi )
        }
\hfill
+
\hfill
 {{3\over
      5}
         \,\Phi^{-3}(\nabla_{i}  V )(\nabla_{i} \nabla^2  V )}
\hfill
-
\hfill
    {{22\over 15}
           \,\Phi^{-4}(\nabla_{i} \Phi )(\nabla_{i} \nabla_{j} \Phi )
        (\nabla_{j} \Phi )}
\nl
-
    {{7\over
      5}
      \,\Phi^{-4}(\nabla_{i}  V )
        (\nabla_{i} \nabla_{j}  V )(\nabla_{j} \Phi )}
\hfill
+
\hfill
    {{11\over
      10}
           \,\Phi^{-5}{{(\nabla \Phi )}^4}}
\hfill
+
\hfill
    {{2\over
      3}
          \,\Phi^{-5}{{(\nabla  V )}^2}{{(\nabla \Phi )}^2}}
\nl
-
    {{11\over
      15}
           \, \Phi^{-4}(\nabla_{i}  V )
        (\nabla_{i} \nabla_{j} \Phi )(\nabla_{j}  V )}
\hfill
+
\hfill
    {{22\over 15}\, \Phi^{-5}
        \left((\nabla_{i} \Phi )(\nabla_{i}  V )\right)^2}
\hfill
-
\hfill
    {{11\over 15}\, \Phi^{-4}
        (\nabla^2 V )(\nabla_{j} \Phi)(\nabla_{j}  V )}
\nl
+
    {{4\over
      15}
         \,\Phi^{-5}{{(\nabla  V )}^4}}
\hfill
-
\hfill
    {{43\over 30}
           \,\Phi^{-4}{{(\nabla \Phi )}^2}\,
        (\nabla^2 \Phi )}
\hfill
-
\hfill
    {{1\over
      3}\,\Phi^{-4}
        {{(\nabla  V )}^2}(\nabla^2 \Phi )}
\nl
+
    {{8\over 15}\,
        \Phi^{-3} (\nabla^2 \Phi )^2}
\hfill
-
\hfill
{{1\over
      30}\,\Phi^{-3}
        (\nabla^2  V )^2}
\Biggl. \Bigg\} \hfill \hfill
\rule[-5mm]{0cm}{1mm}
\\\hline
\rho^{\rm sea}_{E4} =
\frac{\hbar}{24 \, \pi^2} \, \Biggl\{ \Biggr.
-
{{1 \over
      10}
         \, \Phi^{-1}(\nabla^4 \Phi )}
\rule{0cm}{8mm}
\hfill
+
\hfill
{{11 \over 60}
        \, \Phi^{-2} {{(\nabla_{i} \nabla_{j} \Phi )}^2}}
\hfill
+
\hfill
    {{1\over 4}
        \, \Phi^{-2} {{(\nabla_{i} \nabla_{j} V)}^2}}
\nl
+
    {{1 \over
        2}
         \, \Phi^{-2} (\nabla_{i} \Phi )  (\nabla_{i} \nabla^2 \Phi )}
\hfill
+
\hfill
    {{4 \over
      5}
         \, \Phi^{-2} (\nabla_{i} V) (\nabla_{i} \nabla^2 V)}
\hfill
-
\hfill
    {{11\over 15}
         \, \Phi^{-3} (\nabla_{i} \Phi )  (\nabla_{i} \nabla_{j} \Phi )
          (\nabla_{j} \Phi )}
\nl
-
    {{8 \over
      5}
         \, \Phi^{-3} (\nabla_{i} V) (\nabla_{i} \nabla_{j} V) (\nabla_{j} \Phi
)}
\hfill
+
\hfill
    {{11 \over 24}
            \, \Phi^{-4} {{ (\nabla \Phi )}^4}}
\hfill
+
\hfill
    {{5  \over 6}
             \, \Phi^{-4} {{ (\nabla V)}^2}
        {{(\nabla \Phi )}^2}}
\nl
 - {{2 \over
     3}
         \,  \Phi^{-3} (\nabla_{i} V) (\nabla_{i} \nabla_{j} \Phi ) (\nabla_{j}
V)}
\hfill
 +
\hfill
{{49 \over 30}
         \, \Phi^{-4} \left((\nabla_{i} \Phi ) (\nabla_{i} V)\right)^2 }
\hfill \hfill \hfill
\nl
 -
   {{7 \over
     5}
         \,\Phi^{-3} (\nabla^2 V) (\nabla_{i} \Phi ) (\nabla_{i} V)}
\hfill +
\hfill    {{1 \over 5}
          \, \Phi^{-4} {{  (\nabla V)}^4}
        }
\hfill
 -
\hfill
    {{13 \over
      30}
        \, \Phi^{-3} {{  (\nabla \Phi )}^2}  (\nabla^2 \Phi )}
\nl
-
    {{1 \over
      3}
       \, \Phi^{-3} {{ (\nabla V)}^2} (\nabla^2 \Phi )}
\hfill
+
\hfill
  {{1  \over
     15}
       \Phi^{-2}  \, (\nabla^2 \Phi )^2}
\hfill
+
\hfill
 {{7 \over
       20}
         \, \Phi^{-2} (\nabla^2 V)^2}
\Biggl. \Biggr\}
\hfill + \hfill
\rho^{\rm sea}_{4} \, V
\hfill \hfill
\rule[-5mm]{0cm}{1mm}
\\\hline
\end{array}
$$
\\
Table \ref{t-den-sea}
\end{center}
%\end{table}

\newpage
%\begin{table}[hbt]
\begin{center}
$$
\begin{array}{|c|c*{4}{r@{.}l}|}\hline
%
%Woods_C12.dat
^A_Z\rm{X} & & \multicolumn{2}{c}{U_o}
          & \multicolumn{2}{c}{R_o}
          & \multicolumn{2}{c}{a_o}
          & \multicolumn{2}{c|}{\sigma} \\\hline
%
%Woods_O16.dat
^{16}_{8}\rm{O}
    & V:  & 343&8 & 2&630 & 0&5602 & 2&33\\
    &\phi:& 415&4 & 2&624 & 0&6156 & 2&16\\
%
%Woods_Ca40.dat
^{40}_{20}\rm{Ca}
    & V:  & 401&5 & 3&518 & 0&6831 & 5&76\\
    &\phi:& 478&8 & 3&550 & 0&7175 & 5&19\\
%
%Woods_Ca48.dat
^{48}_{20}\rm{Ca}
    & V:  & 386&7 & 3&880 & 0&6074 & 5&35\\
    &\phi:& 464&5 & 3&902 & 0&6458 & 4&52\\
%
%Woods_Ni56.dat
^{56}_{28}\rm{Ni}
    & V:  & 379&2 & 4&175 & 0&5265 & 4&97\\
    &\phi:& 457&9 & 4&190 & 0&5698 & 4&57\\
%
%Woods_Zr90.dat
^{90}_{40}\rm{Zr}
    & V:  & 374&0 & 4&954 & 0&5752 & 5&28\\
    &\phi:& 452&2 & 4&973 & 0&6193 & 5&21\\
%
%Woods_Sn132.dat
^{132}_{50}\rm{Sn}
    & V:  & 371&7 & 5&685 & 0&5745 & 1&86\\
    &\phi:& 449&6 & 5&712 & 0&6151 & 2&88\\
%
%Woods_Pb208.dat
^{208}_{82}\rm{Pb}
    & V:  & 377&0 & 6&634 & 0&6200 & 1&99\\
    &\phi:& 455&0 & 6&669 & 0&6571 & 1&94\\\hline
\end{array}
$$
\\
Table \ref{t-fit}
\end{center}
%\end{table}

\newpage

%\begin{table}[hbt]
\begin{center}
\begin{tabular}{|c|*{2}{r@{.}l|}*{2}{r@{.}l|}*{2}{r@{.}l|}}\cline{2-13}
\multicolumn{1}{c}{ }
        & \multicolumn{4}{|c|}{$\sigma$-$\omega$}
        & \multicolumn{4}{|c|}{WS}
        & \multicolumn{4}{|c|}{Exp.} \\ \hline
$^A_Z$X & \multicolumn{2}{|c|}{$B/A$}
        & \multicolumn{2}{|c|}{m.s.c.r.}
        & \multicolumn{2}{|c|}{$B/A$}
        & \multicolumn{2}{|c|}{m.s.c.r.}
        & \multicolumn{2}{|c|}{$B/A$}
        & \multicolumn{2}{|c|}{m.s.c.r.}\\ \hline
$^{16}_{8}$O    & 3&33   & 2&84\phantom{$^{*}$} &
                  3&38   & 2&83 &
                  7&98   & 2&73\\
$^{40}_{20}$Ca  & 6&27   & 3&48$^{*}$           &
                  6&22   & 3&48 &
                  8&55   & 3&48\\
$^{48}_{20}$Ca  & 6&50   & 3&47\phantom{$^{*}$} &
                  6&34   & 3&49 &
                  8&67   & 3&47\\
$^{56}_{28}$Ni  & 7&23   & 3&72\phantom{$^{*}$} &
                  7&00   & 3&73 &
                  8&64   & \multicolumn{2}{|c|}{}\\
$^{90}_{40}$Zr  & 8&35   & 4&22\phantom{$^{*}$} &
                  8&10   & 4&21 &
                  8&71   & 4&27\\
$^{132}_{50}$Sn & 8&80   & 4&60\phantom{$^{*}$} &
                  8&58   & 4&63 &
                  8&36   & \multicolumn{2}{|c|}{}\\
$^{208}_{82}$Pb & 9&83   & 5&34\phantom{$^{*}$} &
                  9&59   & 5&37 &
                  7&87   & 5&50 \\\hline
\end{tabular}\\[1.5ex]
Table \ref{t-prop}
\end{center}
%\end{table}

\newpage

%\begin{table}[hbt]
\begin{center}
\begin{tabular}{|c|c|*{3}{r}*{2}{|r}|}\cline{3-7}
\multicolumn{1}{c}{} & \multicolumn{1}{c|}{} &
        \multicolumn{1}{c}{WK$_0$} & \multicolumn{1}{c}{WK$_2$} &
\multicolumn{1}{c|}{WK$_4$} & \multicolumn{1}{c|}{WS} &
\multicolumn{1}{c|}{SA} \\\hline
 $^{16}{\rm O}$ & $\pi,\nu$& -143.4\phantom{0} &
-143.9\phantom{5} & -143.4\phantom{0} & -146.6\phantom{0} & \\
 $^{40}{\rm Ca}$ &$\pi,\nu$&
-467.5\phantom{0} & -475.2\phantom{0} & -473.9\phantom{0} &
-476.4\phantom{0}&\\

$^{48}{\rm Ca}$ &$\pi$& -554.1\phantom{0} & -561.2\phantom{0} &
-559.4\phantom{0} & -562.0\phantom{0} & \\
 &$\nu$& -625.2\phantom{0} &
-639.0\phantom{0} &-636.9\phantom{0} & -639.1\phantom{0} & \\
$^{56}{\rm Ni}$ &$\pi,\nu$& -735.1\phantom{0} &
-749.0\phantom{0} & -746.5\phantom{0} & -751.0\phantom{0}
& \\
$^{90}{\rm Zr}$ &$\pi$& -1182\phantom{.00} & -1196\phantom{.00} &
-1193\phantom{.00} & -1193\phantom{.00} & -1193\\
 &$\nu$& -1282\phantom{.00} &
-1303\phantom{.00} &-1300\phantom{.00} & -1309\phantom{.00} & \\
 $^{132}{\rm Sn}$ &$\pi$
&
-1709\phantom{.00}&-1722\phantom{.00}&-1719\phantom{.00}&-1728\phantom{.00}&
-1720\\
 &$\nu$&
-2060\phantom{.00} &-2097\phantom{.00}&-2092\phantom{.00}&
-2106\phantom{.00}& \\
 $^{208}{\rm
Pb}$ &$\pi$& -2848\phantom{.00} & -2868\phantom{.00}&-2864\phantom{.00}&
-2878\phantom{.00}
&-2866\\
 &$\nu$& -3346\phantom{.00} & -3395\phantom{.00}
&-3391\phantom{.00}&-3407\phantom{.00}& \\\hline
\end{tabular}\\[1.5ex]
Table \ref{t-WS-res}
\end{center}
%\end{table}

\newpage

%\begin{table}[ht]
\begin{center}
\footnotesize
\[
\begin{array}{||rlrlrl||}\hline
\multicolumn{6}{||c||}{W_2(x,k)= \frac{k^2}{D}
\mbox{tr[ ]}}\\ \hline
& & & & &\\
\multicolumn{6}{||c||}{\Delta_{\mu}^2}\\
& & & & &\\
\hline \hline
\multicolumn{6}{||c||}{W_4(x,k)=-2 \frac{k^4}{D(D+2)}
\mbox{tr[ ]}}\\ \hline
& & & & &\\
  -2& \Delta_{\mu}^4
& + & (\Delta_{\mu} \Delta_{\nu})^2
&  +2 & (\Delta \Delta_{\mu \mu})^2\\
- & (F_{\mu \nu} \Delta^2)^2
& -4i & \Delta F_{\mu \nu} \Delta \Delta_{\mu} \Delta_{\nu}
& & \\
& & & & &\\
\hline \hline
\multicolumn{6}{||c||}{W_6(x,k)=32 \frac{k^6}{D(D+2)(D+4)}
 \mbox{tr[ ]}}\\ \hline
& & & & &\\
    \frac{5}{3} & \Delta_{\mu }^6
&   -\frac{1}{2} & (\Delta_{\mu } \Delta_{\nu } \Delta_{\mu })^2
&   + \frac{1}{6} & (\Delta_{\mu } \Delta_{\nu } \Delta_{\rho })^2\\
  + \frac{1}{2} & (\Delta_{\mu }^2 \Delta_{\nu })^2
&    - & \Delta_{\mu }^2 (\Delta_{\nu } \Delta_{\rho })^2
&   -4 & \Delta \Delta_{\mu } \Delta_{\nu } \Delta_{\mu } \Delta_{\rho}
 \Delta_{\rho \nu }\\
    + 4 & \Delta \Delta_{\mu } \Delta_{\nu }^2 \Delta_{\rho }
\Delta_{\rho \mu }
& + \frac{3}{2} & \Delta^2 \Delta_{\mu \mu } \Delta_{\nu }^2
\Delta_{\rho \rho }
&    -4 &  (\Delta \Delta_{\mu \mu } )^2  \Delta_{\nu }^2 \\
   -2 & (\Delta \Delta_{\mu } \Delta_{\nu \mu } )^2
&    -4 & \Delta \Delta_{\mu } \Delta_{\nu \mu } \Delta_{\nu } \Delta
\Delta_{\rho \rho }
& -2 & \Delta \Delta_{\mu } \Delta_{\nu } \Delta \Delta_{\rho \mu }
\Delta_{\rho \nu }   \\
  + 2 & \Delta \Delta_{\mu } \Delta_{\nu } \Delta \Delta_{\rho \nu }
 \Delta_{\rho \mu }
& + \frac{5}{2} & (\Delta \Delta_{\mu } \Delta_{\nu \nu } )^2
& - & \Delta \Delta_{\mu } \Delta_{\nu \nu } \Delta \Delta_{\rho \rho}
 \Delta_{\mu }\\
  +2 & \Delta \Delta_{\mu } \Delta_{\nu \nu } \Delta_{\mu } \Delta
\Delta_{\rho \rho }
&   - \frac{2}{3} & (\Delta \Delta_{\mu \mu } )^3
& +3 & \Delta^2 \Delta_{\mu \mu } \Delta_{\nu } \Delta
\Delta_{\rho \rho \nu }   \\
  +\frac{3}{4} &  (\Delta^2 \Delta_{\nu \nu \mu } )^2
& & & & \\
& & & & & \\
   +2 \, i & F_{\mu \nu} \Delta \Delta_{\mu } \Delta_{\nu }
 \Delta_{\rho }^2  \Delta
& +2 \, i & F_{\mu \nu} \Delta \Delta_{\mu } \Delta_{\rho }
 \Delta_{\nu } \Delta_{\rho } \Delta
&  +i & F_{\mu \nu} \Delta \Delta_{\mu } \Delta_{\rho }^2
\Delta_{\nu } \Delta \\
  - i & F_{\mu \nu} \Delta \Delta_{\rho } \Delta_{\mu }
\Delta_{\nu } \Delta_{\rho } \Delta
&   +2 \, i & F_{\mu \nu} \Delta_{\rho } \Delta \Delta_{\mu }
 \Delta_{\nu } \Delta \Delta_{\rho }
& +4 \, i & F_{\mu \nu} \Delta^2 \Delta_{\mu } \Delta_{\rho \rho }
 \Delta_{\nu } \Delta   \\
  +4 \, i & F_{\mu \nu} \Delta^2 \Delta_{\rho \mu } \Delta_{\rho }
 \Delta_{\nu } \Delta
& +4 \, i & F_{\mu \nu} \Delta^2 \Delta_{\rho } \Delta_{\mu \rho }
 \Delta \Delta_{\nu }
&  +4 \, i & F_{\mu \nu} \Delta^2
 \Delta_{\rho \rho } \Delta \Delta_{\mu }
 \Delta_{\nu }\\
 +4 \, i & F_{\mu \nu} \Delta \Delta_{\mu } \Delta \Delta_{\rho \rho }
\Delta_{\nu } \Delta
& +2 \, i & F_{\mu \nu} \Delta^2 \Delta_{\rho \mu }
\Delta_{\rho \nu } \Delta^2
& +4 \, i & F_{\mu \nu} \Delta^2 \Delta_{\mu } \Delta_{\rho \rho \nu }
 \Delta^2   \\
  +4 \, i & F_{\mu \nu} \Delta^2 \Delta_{\rho \rho \mu } \Delta
\Delta_{\nu } \Delta
& +2 \, i & \Delta^2 F_{\mu \nu \nu } \Delta^2 \Delta_{\mu }
\Delta_{\rho \rho }
& & \\
& & & & & \\
  +\frac{1}{2} & (\Delta F_{\mu \nu} \Delta)^2 \Delta_{\rho }^2
&+& F_{\mu \nu} \Delta^2 F_{\mu \rho} \Delta \Delta_{\nu} \Delta_{\rho}
 \Delta
&-&F_{\mu \nu} \Delta^2 F_{\mu \rho} \Delta \Delta_{\rho} \Delta_{\nu}
 \Delta \\
 -4&F_{\mu \nu} \Delta^3 F_{\mu \rho} \Delta_{\rho} \Delta \Delta_{\nu}
&   +  & F_{\mu \nu} \Delta^2 \Delta_{\rho } F_{\mu \nu} \Delta_{\rho }
\Delta^2
& -2 & F_{\mu \nu} \Delta^2 \Delta_{\rho } F_{\mu \rho} \Delta^2
\Delta_{\nu }   \\
   -  & (\Delta F_{\mu \nu} \Delta \Delta_{\mu })^2
& -\frac{1}{4} &(\Delta F_{\mu \nu} \Delta \Delta_{\rho })^2
&+ 2 & F_{\mu \nu} \Delta^3 F_{\mu \nu} \Delta^2 \Delta_{\rho \rho }\\
  -4 & F_{\mu \nu} \Delta^3 F_{\mu \rho \rho } \Delta^2 \Delta_{\nu }
&   - \frac{1}{2} & (\Delta^3  F_{\mu \nu \nu } )^2
& & \\
& & & & &\\
-\frac{1}{3}\,i & F_{\mu \nu}
 \Delta^2 F_{\mu \rho} \Delta^2 F_{\nu \rho}
 \Delta^2
& & & &\\
& & & & &\\  \hline
\end{array}
\]
\\
Table \ref{t-Chan}
\end{center}
%\end{table}

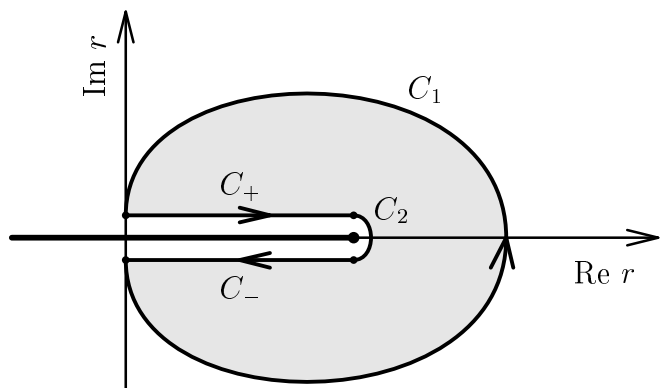
\end{document}